\documentclass[
jmp,
amsmath,amssymb,
reprint,%
]{revtex4-1}

\usepackage{graphicx}
\usepackage{dcolumn}
\usepackage{bm}

\usepackage[utf8]{inputenc}
\usepackage[T1]{fontenc}
\usepackage{mathptmx}
\usepackage{etoolbox}

\usepackage{amsmath,amsfonts,amssymb,caption,color,epsfig,graphics,graphicx,latexsym,mathrsfs,revsymb,theorem,url,verbatim,epstopdf}

\usepackage{hyperref}

\hypersetup{colorlinks,linkcolor={blue},citecolor={blue},urlcolor={red}} 

\usepackage{overpic}

\allowdisplaybreaks

\usepackage{hyperref}

\newtheorem{definition}{Definition}
\newtheorem{proposition}[definition]{Proposition}
\newtheorem{lemma}[definition]{Lemma}

\newtheorem{theorem}[definition]{Theorem}
\newtheorem{corollary}[definition]{Corollary}
\newtheorem{conjecture}[definition]{Conjecture}

\newtheorem{remark}[definition]{Remark}
\newtheorem{example}[definition]{Example}
\newtheorem{question}[definition]{Question}

\def\squareforqed{\hbox{\rlap{$\sqcap$}$\sqcup$}}
\def\qed{\ifmmode\squareforqed\else{\unskip\nobreak\hfil
		\penalty50\hskip1em\null\nobreak\hfil\squareforqed
		\parfillskip=0pt\finalhyphendemerits=0\endgraf}\fi}
\def\endenv{\ifmmode\;\else{\unskip\nobreak\hfil
		\penalty50\hskip1em\null\nobreak\hfil\;
		\parfillskip=0pt\finalhyphendemerits=0\endgraf}\fi}
\newenvironment{proof}{\noindent \textbf{{Proof.~} }}{\qed}
\def\Dbar{\leavevmode\lower.6ex\hbox to 0pt
	{\hskip-.23ex\accent"16\hss}D}
\makeatletter
\def\url@leostyle{%
	\@ifundefined{selectfont}{\def\UrlFont{\sf}}{\def\UrlFont{\small\ttfamily}}}
\makeatother
\def\bcj{\begin{conjecture}}
	\def\ecj{\end{conjecture}}
\def\bcr{\begin{corollary}}
	\def\ecr{\end{corollary}}
\def\bd{\begin{definition}}
	\def\ed{\end{definition}}
\def\bea{\begin{eqnarray}}
	\def\eea{\end{eqnarray}}
\def\bem{\begin{enumerate}}
	\def\eem{\end{enumerate}}
\def\bex{\begin{example}}
	\def\eex{\end{example}}
\def\bim{\begin{itemize}}
	\def\eim{\end{itemize}}
\def\bl{\begin{lemma}}
	\def\el{\end{lemma}}
\def\bma{\begin{bmatrix}}
	\def\ema{\end{bmatrix}}
\def\bpf{\begin{proof}}
	\def\epf{\end{proof}}
\def\bpp{\begin{proposition}}
	\def\epp{\end{proposition}}
\def\bqu{\begin{question}}
	\def\equ{\end{question}}
\def\br{\begin{remark}}
	\def\er{\end{remark}}
\def\bt{\begin{theorem}}
	\def\et{\end{theorem}}

\def\btb{\begin{tabular}}
	\def\etb{\end{tabular}}

\newcommand{\nc}{\newcommand}

\def\a{\alpha}

\def\p{\pi}

\nc{\bbA}{\mathbb{A}} \nc{\bbB}{\mathbb{B}} \nc{\bbC}{\mathbb{C}}
\nc{\bbD}{\mathbb{D}} \nc{\bbE}{\mathbb{E}} \nc{\bbF}{\mathbb{F}}
\nc{\bbG}{\mathbb{G}} \nc{\bbH}{\mathbb{H}} \nc{\bbI}{\mathbb{I}}
\nc{\bbJ}{\mathbb{J}} \nc{\bbK}{\mathbb{K}} \nc{\bbL}{\mathbb{L}}
\nc{\bbM}{\mathbb{M}} \nc{\bbN}{\mathbb{N}} \nc{\bbO}{\mathbb{O}}
\nc{\bbP}{\mathbb{P}} \nc{\bbQ}{\mathbb{Q}} \nc{\bbR}{\mathbb{R}}
\nc{\bbS}{\mathbb{S}} \nc{\bbT}{\mathbb{T}} \nc{\bbU}{\mathbb{U}}
\nc{\bbV}{\mathbb{V}} \nc{\bbW}{\mathbb{W}} \nc{\bbX}{\mathbb{X}}
\nc{\bbZ}{\mathbb{Z}}

\nc{\bA}{{\bf A}} \nc{\bB}{{\bf B}} \nc{\bC}{{\bf C}}
\nc{\bD}{{\bf D}} \nc{\bE}{{\bf E}} \nc{\bF}{{\bf F}}
\nc{\bG}{{\bf G}} \nc{\bH}{{\bf H}} \nc{\bI}{{\bf I}}
\nc{\bJ}{{\bf J}} \nc{\bK}{{\bf K}} \nc{\bL}{{\bf L}}
\nc{\bM}{{\bf M}} \nc{\bN}{{\bf N}} \nc{\bO}{{\bf O}}
\nc{\bP}{{\bf P}} \nc{\bQ}{{\bf Q}} \nc{\bR}{{\bf R}}
\nc{\bS}{{\bf S}} \nc{\bT}{{\bf T}} \nc{\bU}{{\bf U}}
\nc{\bV}{{\bf V}} \nc{\bW}{{\bf W}} \nc{\bX}{{\bf X}}
\nc{\bZ}{{\bf Z}}

\nc{\cA}{{\cal A}} \nc{\cB}{{\cal B}} \nc{\cC}{{\cal C}}
\nc{\cD}{{\cal D}} \nc{\cE}{{\cal E}} \nc{\cF}{{\cal F}}
\nc{\cG}{{\cal G}} \nc{\cH}{{\cal H}} \nc{\cI}{{\cal I}}
\nc{\cJ}{{\cal J}} \nc{\cK}{{\cal K}} \nc{\cL}{{\cal L}}
\nc{\cM}{{\cal M}} \nc{\cN}{{\cal N}} \nc{\cO}{{\cal O}}
\nc{\cP}{{\cal P}} \nc{\cQ}{{\cal Q}} \nc{\cR}{{\cal R}}
\nc{\cS}{{\cal S}} \nc{\cT}{{\cal T}} \nc{\cU}{{\cal U}}
\nc{\cV}{{\cal V}} \nc{\cW}{{\cal W}} \nc{\cX}{{\cal X}}
\nc{\cZ}{{\cal Z}}

\nc{\hA}{{\hat{A}}} \nc{\hB}{{\hat{B}}} \nc{\hC}{{\hat{C}}}
\nc{\hD}{{\hat{D}}} \nc{\hE}{{\hat{E}}} \nc{\hF}{{\hat{F}}}
\nc{\hG}{{\hat{G}}} \nc{\hH}{{\hat{H}}} \nc{\hI}{{\hat{I}}}
\nc{\hJ}{{\hat{J}}} \nc{\hK}{{\hat{K}}} \nc{\hL}{{\hat{L}}}
\nc{\hM}{{\hat{M}}} \nc{\hN}{{\hat{N}}} \nc{\hO}{{\hat{O}}}
\nc{\hP}{{\hat{P}}} \nc{\hR}{{\hat{R}}} \nc{\hS}{{\hat{S}}}
\nc{\hT}{{\hat{T}}} \nc{\hU}{{\hat{U}}} \nc{\hV}{{\hat{V}}}
\nc{\hW}{{\hat{W}}} \nc{\hX}{{\hat{X}}} \nc{\hZ}{{\hat{Z}}}

\nc{\hn}{{\hat{n}}}

\def\diag{\mathop{\rm diag}}
\def\dim{\mathop{\rm dim}}

\def\lin{\mathop{\rm span}}

\def\max{\mathop{\rm max}}

\def\tr{\mathop{\rm Tr}}

\def\lra{\leftrightarrow}

\newcommand{\bra}[1]{\langle#1|}
\newcommand{\ket}[1]{|#1\rangle}

\newcommand{\ketbra}[2]{|#1\rangle\!\langle#2|}

\begin{document}
	
	\preprint{AIP/123-QED}
	
	\title[Unextendible product operator basis]{Unextendible product operator basis}
	\author{Mengyao Hu}\email{mengyao@lorentz.leidenuniv.nl}
	\affiliation{School of Mathematical Sciences, Beihang University, Beijing 100191, China}
	\affiliation{Instituut-Lorentz, Universiteit Leiden, P.O. Box 9506, 2300 RA Leiden, The Netherlands}

	\author{Lin Chen}\email{linchen@buaa.edu.cn (corresponding author)}
	\affiliation{School of Mathematical Sciences, Beihang University, Beijing 100191, China}
	\affiliation{LMIB(Beihang University), Ministry of Education, China}
	\affiliation{International Research Institute for Multidisciplinary Science, Beihang University, Beijing 100191, China}

	\author{Fei Shi}\email{shifei@mail.ustc.edu.cn}
	\affiliation{Department of Computer Science, The University of Hong Kong, Pokfulam Road, Hong Kong, China}
	\affiliation{School of Cyber Security, University of Science and Technology of China, Hefei 230026, China}
	
	\author{Xiande Zhang}\email{drzhangx@ustc.edu.cn}
	\affiliation{School of Mathematical Sciences, University of Science and Technology of China, Hefei 230026, China}
	\affiliation{Hefei National Laboratory, University of Science and Technology of China, Hefei 230088, China}
	
	\author{Jordi Tura}\email{tura@lorentz.leidenuniv.nl}
	\affiliation{Instituut-Lorentz, Universiteit Leiden, P.O. Box 9506, 2300 RA Leiden, The Netherlands}
	\date{\today}
	
	\begin{abstract}
		Quantum nonlocality is associated with the local indistinguishability of orthogonal states. Unextendible product basis (UPB), a widely used tool in quantum information, exhibits nonlocality which is the powerful resource for quantum information processing. In this work we extend the definitions of nonlocality and genuine nonlocality from states to operators. We also extend UPB to the notions of  unextendible product operator basis, unextendible product unitary operator basis (UPUOB) and strongly UPUOB. We construct their examples, and show the nonlocality of some strongly UPUOBs under local operations and classical communications. We study the phenomenon of these operators acting on quantum states. As an application, we distinguish the two-dimensional strongly UPUOB which only consumes three ebits of entanglement. Our results imply that such UPUOBs exhibit nonlocality as UPBs and the distinguishability of them requires entanglement resources. 
	\end{abstract}
	
	\maketitle
	
	\section{Introduction}
	\label{sec:int} 
	If a set of orthogonal quantum states cannot be distinguished by local operations and classical communications (LOCC), it is locally indistinguishable. Such a set exhibits the phenomenon of quantum nonlocality \cite{PhysRevA.102.042202,PhysRevLett.122.040403}. In many previous works, authors considered the problem of distinguishing a given set of states by LOCC \cite{PhysRevA.59.1070, PhysRevLett.120.240402, PhysRevLett.113.130401, PhysRevA.77.012304}. For example, unextendible product bases (UPBs) are locally indistinguishable \cite{ISI:000184088700001, PhysRevA.70.022309}. As far as we know, little is known about the local indistinguishability of unitary operators \cite{lanyon2009simplifying, procopio2015experimental, bagchi2016uncertainty}. The locality of $n$-partite unitary operators means that they are the tensor products of unitary operators locally acting on subsystems, that is, $U_{A_1}\otimes \cdots \otimes U_{A_j} \otimes \cdots \otimes U_{A_n}$, where the $j$-th system performs the operator $U_{A_j}$ \cite{PhysRevA.89.062326, PhysRevA.62.052317}. So there is no entanglement between distributed parties. The local unitary operators can be perfectly implemented by LOCC. However, there remains whether the discrimination of $n$-partite ($n\geq 2$) product unitary operators can be done locally or in some $m$-partitions ($m \in \{2,\cdots,n-1\}$) unkown. Our work builds towards filling this gap. 
	
	Quantum nonlocality is a non-additive resource which can be activated \cite{cite-key}. Recently, quantum nonlocality was extended to genuine quantum nonlocality \cite{RevModPhys.86.419, PhysRevLett.123.140401} and strong nonlocality \cite{PhysRevLett.122.040403}. They play a key role in state discrimination and multipartite secret sharing \cite{PhysRevLett.126.210505}. 
	It has been proved that any discrete finite-dimensional unitary operators can be constructed in the laboratory using optical devices \cite{reck1994experimental}. A series of them are the bases of quantum computation or algorithms \cite{PhysRevA.96.032321, 8049724, song2019quantum}. The product unitary operators, such as product Pauli operators, are useful in quantum information masking under noise \cite{hu2021genuine}. For maximally entangled states(MESs), the local distinguishability of them has been well studied \cite{PhysRevA.91.052314, PhysRevLett.109.020506, PhysRevLett.92.177905}. It has been proved that any $l$ mutually orthogonal generalized $d \times d$ Bell states are locally distinguishable if $d$ is a prime and $l(l-1)<2d$ \cite{PhysRevLett.92.177905}. We extend the distinguishability of orthogonal MESs to the nonlocality of unitary operators in terms of MESs. That is, if some orthogonal MESs can be distinguished by LOCC, are the tensor powers of these orthogonal MESs distinguishable by LOCC? Our work gives the answer for the tensor of bipartite two-level MESs.

	\begin{figure}[!h] 
		\center{\includegraphics[width=8.5cm]  {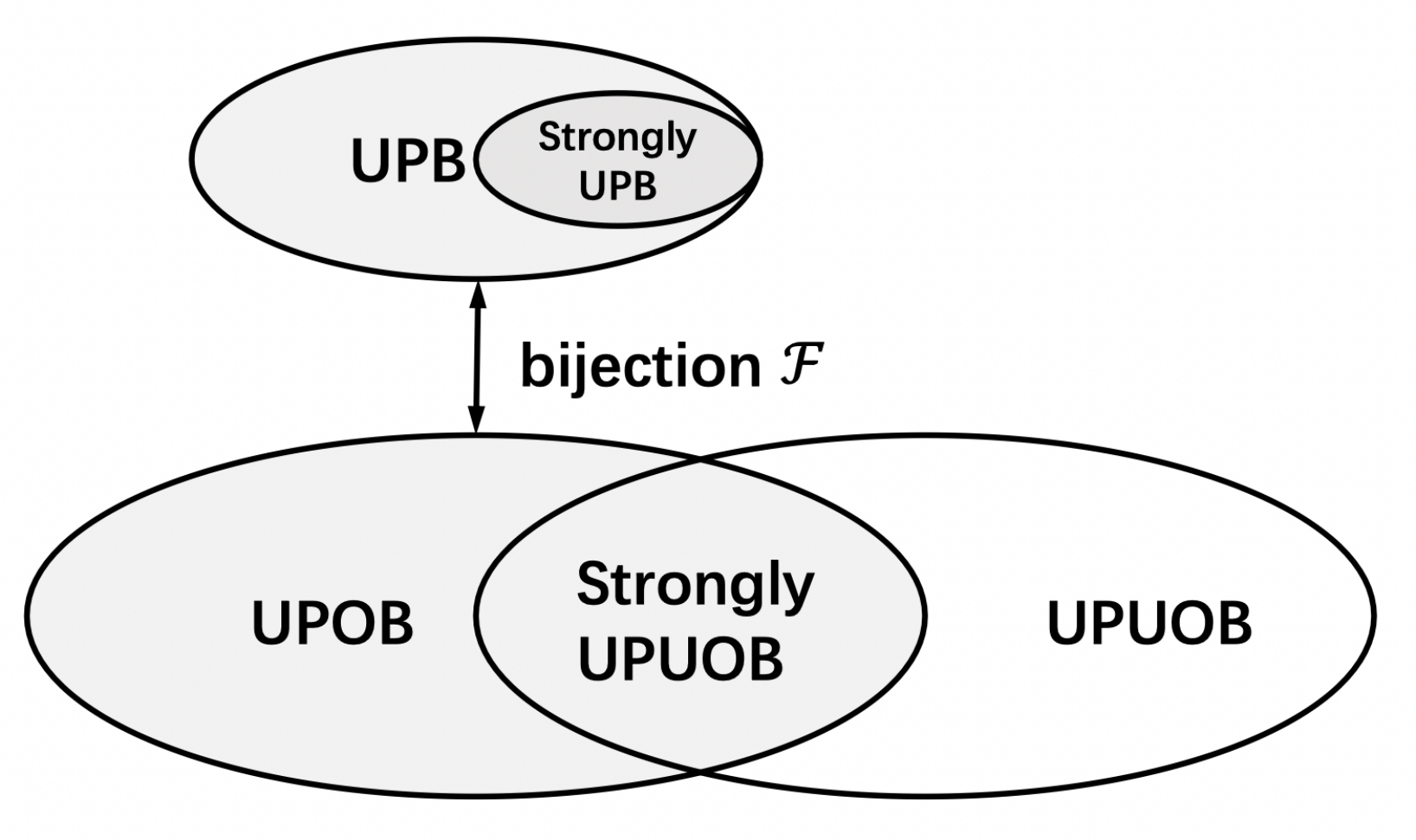}} 
		\caption{\label{fig:isomorphism} Using the bijection $\cF$ from vectors to matrices in \eqref{eq:bijection_general}, one can obtain a UPOB from a UPB and a strongly UPUOB from strongly UPB, respectively. Further, the strongly UPUOBs are the intersections of UPUOBs and UPOBs. }   
	\end{figure}
	
	In this work, motivated by the nonlocality and genuine nonlocality of orthogonal product states in Definition \ref{def:nonlocalstate}, we define the counter part of unitary operators in Definition \ref{def:nonlocaloperator}. In Definition \ref{def:UPO}, extended from  UPBs, we propose the notion of unextendible product operator basis (UPOB), unextendible product unitary operator basis (UPUOB) and strongly UPUOB. The connection among them is shown in FIG. \ref{fig:isomorphism}. Then we give the properties of UPOB and UPUOB in Lemmas \ref{le:UUO_qd}-\ref{le:UPUOB_qd}. By using Lemma \ref{le:local_rank}, we prove that arbitrary tensor powers of bipartite UPOBs are again UPOBs. Next, a bijection from vectors to matrices is proposed to construct UPOBs from UPBs. In Proposition \ref{pro:UPOB_size}, we show the existence of UPOBs. We also show that there exists an $n$-qubit strongly UPUOB for $n \geq 2$ in Theorem \ref{thm:UPUOB2n}. In Example \ref{ex:UPUOB23} and Theorem \ref{thm:UPUOBqd}, we prove that some  bipartite UPUOBs are not strongly UPUOBs. In Theorem \ref{thm:UPUOB22nonlocal}, we show that the two-qubit strongly UPUOB is genuinely nonlocal in terms of the two-level MESs. Finally in Theorem \ref{thm:ebitent}, we  prove that the strongly UPUOB can be distinguished by using three ebits of entanglement. It can be used in multiparty secret sharing and unitary gates discrimination. 
	
	The remaining of this paper is organized as follows. In Sec. \ref{sec:pre}, we introduce the preliminary knowledge used in this paper, such as the nonlocality, UPBs and UPOBs.  In Sec. \ref{sec:construction} we construct examples of UPOBs, UPUOBs and strongly UPUOBs. Then we study the nonlocality of UPUOBs and apply our results in Sec. \ref{sec:app}. We conclude in Sec. \ref{sec:con}.

	\section{Preliminaries}
	\label{sec:pre}
	
	In this section, we introduce the notations and preliminary facts used in this paper. We may not normalize states and operators for simplicity. Let $\bbC^d$ be the $d$-dimensional Hilbert space. Assume that $\{\ket{i}_A\}_{i=0}^{d_A-1}$ and $\{\ket{j}_B\}_{j=0}^{d_B-1}$ are computational bases in $\cH_A$ and $\cH_B$ respectively, where $\dim(\cH_A)=d_A$, $\dim(\cH_B)=d_B$. For any linear operator $\cM$ from $\cH_A$ to $\cH_B$, we assume that the $d_B\times d_A$ matrix $M$ is the matrix representation of the operator $\cM$ under the computational bases  $\{\ket{i}_A\}_{i=0}^{d_A-1}$ and $\{\ket{j}_B\}_{j=0}^{d_B-1}$. In general, we do not distinguish the operator $\cM$ and the  matrix $M$. We shall use the notation $\mathbb{O}(\cH_A,\cH_B)$ for the space of linear operators from $\cH_A$ to $\cH_B$, and $\bbM_{d_B,d_A}$ for the space of $d_B\times d_A$ matrices.
	Obviously, $\mathbb{O}(\cH_A,\cH_B)\cong \bbM_{d_B,d_A}$.
	We also do not distinguish the space $\mathbb{O}(\cH_A,\cH_B)$ and the space $\bbM_{d_B,d_A}$. Specially, we denote $\mathbb{O}(\cH)=\mathbb{O}(\cH,\cH)$.  For any two matrices $M_1, M_2\in \mathbb{O}(\cH_A,\cH_B)$, the inner product of $M_1$ and $M_2$ is $\tr{(M_1^{\dagger}M_2)}$.  Further, two subspaces $\cO, \cO^\perp$ of $\mathbb{O}(\cH_1,\cH_2)$ are said to be complementary if their direct sum gives the entire space, and every matrix in $\cO$ is orthogonal to every matrix in $\cO^\perp$.  For a product matrix $M=M_A\otimes M_B \in \cO(\cH_{A})\otimes \cO(\cH_{B})$, we shall refer to $M_A$ as the $A$ partition of $M$, and $M_B$ as the $B$ partition of $M$. For a set $\mathbb{U}$, we denote the cardinality of it by $|\mathbb{U}|$. We say that the linearly independent operator set $\{U_i,i=1,\cdots,d^2\}$ is complete if it spans $\mathbb{O}(\cH)$ with $\dim(\cH)=d$. So it is a basis. 
	Let $\omega_k=e^{\frac{2\pi i}{k}}$ be a primitive $k$-th root of unity. 
	We say that $M$ and $M'$ are equivalent modulo a global phase if and only if there is a real $\theta$ such that the matrix $M'=e^{i\theta}M$. 
	The global phase means there is a real $\theta$ such that the matrix $M$ becomes $e^{i\theta}M$. We denote the Pauli matrices as $\sigma_\alpha$, where $\alpha=x,y,z$. 
	
	Next, we review the orthonormality and  $k$-orthonormality of matrices. We say that $s$ $d\times d$ matrices $U_1,\cdots,U_s$ are orthonormal if $\tr (U_j^\dag U_k)=d\delta_{jk}$ for any $j,k$. Moreover, if $U_1,\cdots,U_s$ are $n$-partite tensor product unitary matrices of system $A_1,\cdots,A_n$ and they are orthonormal, then we refer to them as orthogonal product (OP) unitary operators of system $A_1,\cdots,A_n$. To characterize them, we propose the definition of $k$-orthonormality.
	
	\begin{definition}
		Two $n$-partite OP unitary matrices $\bigotimes^n_{i=1}U_i$ and $\bigotimes^n_{i=1}U_i'$ are called $k$-orthonormal if $U_k$ and $U'_k$ are orthonormal for the smallest $k\in \{1,\cdots,n\}$.
	\end{definition}
	
	The remaining of this section is divided into two subsections.
	In Sec. \ref{subsec:nonlocality}, we introduce the definition of nonlocality and genuine nonlocality of states and operators in Definitions \ref{def:nonlocalstate} and \ref{def:nonlocaloperator}, respectively. In Sec. \ref{subsec:upob}, we introduce the definition of unextendible product operator basis (UPOB), unextendible product unitary operator basis (UPUOB) and strongly UPUOB in Definition \ref{def:UPO}. The connection among them is given in FIG. \ref{fig:isomorphism}. We give some constructions of UPOBs and UPUOBs in Lemmas \ref{le:UUO_qd}-\ref{le:local_rank} and Theorem \ref{thm:UPUOBtensor}.
	
	\subsection{nonlocality}
	\label{subsec:nonlocality}
	
	A set of orthogonal quantum states is locally indistinguishable if it cannot be distinguished by LOCC.  Nonlocality and genuine nonlocality play a key role in various quantum information tasks such as quantum data hiding \cite{matthews2009distinguishability, divincenzo2001quantum} and secret sharing \cite{PhysRevA.78.042309}. In the following, Definition \ref{def:nonlocalstate} reviews the nonlocality and genuine nonlocality. They will be extended to operators in Definition \ref{def:nonlocaloperator}. This will be used in Theorem \ref{thm:UPUOB22nonlocal} to show that the two-qubit strongly UPUOB is genuinely nonlocal in terms of the two-level MES.
	
	\begin{definition}
		\label{def:nonlocalstate}
		Consider the quantum systems $A_1,\cdots,A_n$ in the $n$-partite Hilbert space $\bigotimes^n_{i=1}\mathbb{C}^{d_i}$. 
		
		(i) A set of $n$-partite OP states $\mathbb{B}_{\mathrm{nl}}\equiv \{\ket{\psi_j}=\bigotimes_{i=1}^n\ket{\a^i_j}_{A_i} \mid j=1,2,\cdots,N \leq \prod_{i=1}^nd_i\} \subset \bigotimes_{i=1}^n \mathbb{C}^{d_i}$ is called nonlocal if the states in  $\mathbb{B}_{\mathrm{nl}}$ are locally indistinguishable. 
		
		(ii) A set of $n$-partite OP states $\mathbb{B}_{\mathrm{gnl}}\equiv \{\ket{\psi_j}=\bigotimes_{i=1}^n\ket{\a^i_j}_{A_i} \mid j=1,2,\cdots,N \leq \prod_{i=1}^nd_i\} \subset \bigotimes_{i=1}^n \mathbb{C}^{d_i}$ is called genuinely nonlocal if the states in  $\mathbb{B}_{\mathrm{gnl}}$ are locally indistinguishable for every bipartition of the subsystems.
	\end{definition}
	
	To extend the definition, we define the nonlocality and genuine nonlocality of unitary operators. Most of the related results are about the nonlocality of quantum states and little is known about the operators. So we introduce the nonlocality of operators and study the phenomenon of unitary operators in terms of MESs.
	
	\begin{definition}
		\label{def:nonlocaloperator}
		For $n\geq 2$, consider the quantum systems $A_1,\cdots,A_n$ in the $n$-partite Hilbert space $\bigotimes^n_{i=1}\mathbb{C}^{d_i}$ and $B_1,\cdots,B_n$ in the same Hilbert space. Let 
		\begin{eqnarray}
			\label{eq:a_k}
			\{\ket{a_k}=U_k\otimes I_{B_1,\cdots,B_n} (\ket{\psi}_{A_1B_1} \otimes \cdots \\ \notag \otimes \ket{\psi}_{A_nB_n}) 
			\mid k=1,\cdots,N\leq \prod_{i=1}^nd_i^2\},
		\end{eqnarray}
		where $U_k$ are $n$-partite unitary operators of system $A_1,\cdots,A_n$. Let $\mathbb{U}\equiv \{U_k |k=1,2,\cdots,N\}$.
		
		(i) If the set of states $\{\ket{a_k}\}_{k=1}^N$ is nonlocal, then the set of unitary operators $\mathbb{U}$ is called nonlocal in terms of $\ket{\psi}$. 
		
		(ii) If the set of states $\{\ket{a_k}\}_{k=1}^N$ is genuinely nonlocal, then the set  of unitary operators $\mathbb{U}$ is called genuinely nonlocal in terms of $\ket{\psi}$.
		
		(iii) If the states $\{\ket{a_k}\}_{k=1}^N$  satisfy the conditions in  (i) and (ii) for any bipartite state $\ket{\psi}$, respectively, then the set of unitary operators $\mathbb{U}$ in (i) and (ii) is called nonlocal and genuinely nonlocal, respectively. 
	\end{definition}
	
	Evidently the genuine nonlocality of operators ensures the nonlocality of operators.  So we have extended the nonlocality from states to operators, which will be used in the discrimination of operators in Theorem \ref{thm:UPUOB22nonlocal} and Theorem \ref{thm:ebitent}.
	
	\subsection{UPOB and UPUOB}
	\label{subsec:upob}

	An unextendible product basis (UPB) is a basis of orthonormal product states in the subspace $ \cH \subseteq \bigotimes^n_{i=1}\cH_{i}$ whose complementary subspace $\cH^\perp \subseteq \bigotimes^n_{i=1}\cH_{i}$ does not contain any product state \cite{PhysRevLett.82.5385}. It has been proven that any UPB is locally indistinguishable \cite{PhysRevA.70.022309}. The following is an example of UPB in $\mathbb{C}^4 \otimes \mathbb{C}^4$.
	
	\begin{example}
		\label{ex:UPB}
		The basis $\{\ket{\psi_j}\}_{j=1}^{11}$ is a UPB in $\mathbb{C}^4 \otimes \mathbb{C}^4$ \cite{PhysRevA.101.062329}, where
		\begin{widetext}
			\begin{eqnarray}
				\notag
				\begin{aligned}
					&\ket{\psi_1}=\frac{1}{\sqrt{2}}\ket{0}(\ket{0}-\ket{1}), &&\ket{\psi_2}=\frac{1}{\sqrt{2}}(\ket{0}-\ket{3})\ket{2}, \\
					&\ket{\psi_3}=\frac{1}{\sqrt{3}}(\ket{0}+\omega_3\ket{1}+\omega_3^2\ket{2})\ket{3}, 
					&&\ket{\psi_4}=\frac{1}{\sqrt{3}}(\ket{0}+\omega_3^2\ket{1}+\omega_3\ket{2})\ket{3}, \\
					& \ket{\psi_5}=\frac{1}{\sqrt{3}}(\ket{1}+\omega_3\ket{2}+\omega_3^2\ket{3})\ket{1},
					&&\ket{\psi_6}=\frac{1}{\sqrt{3}}(\ket{1}+\omega_3^2\ket{2}+\omega_3\ket{3})\ket{1}, \\
					& \ket{\psi_7}=\frac{1}{\sqrt{2}}\ket{3}(\ket{0}-\ket{3}),
					&& \ket{\psi_8}=\frac{1}{2}(\ket{1}+\ket{2})(\ket{0}-\ket{2}),\\
					& \ket{\psi_9}=\frac{1}{2}(\ket{1}-\ket{2})(\ket{0}+\ket{2}),	
					&& \ket{\psi_{10}}=\frac{1}{2}(\ket{1}-\ket{2})(\ket{0}-\ket{2}),\\
					&\ket{\psi_{11}}=\frac{1}{4}(\ket{0}+\ket{1}+\ket{2}+\ket{3})(\ket{0}+\ket{1}+\ket{2}+\ket{3}). 
				\end{aligned} 
			\end{eqnarray}
		\end{widetext}
	\end{example}
	
	
	Next, we extend the notion of UPBs to UPOBs. We shall omit the trivial case of complete UPOBs, just like the case of UPBs. 
	
	\begin{definition}
		\label{def:UPO} Let the subspace $\cH =\bigotimes^n_{i=1}\cH_{i}$. Consider the operator subspace $\cO \subset \mathbb{O}(\cH)$ and its complementary subspace $\cO^\perp$.
		
		(i) A set of $n$-partite OP operators is a UPOB in $\cO$ whose complementary subspace $\cO^\perp$ does not contain any product operator. 
		
		(ii) A set of $n$-partite OP unitary operators is a UPUOB in $\cO$ whose complementary subspace $\cO^\perp$ does not contain any product unitary operator. 
		
		(iii) A set of $n$-partite OP unitary operators is a strongly UPUOB in $\cO$ whose complementary subspace $\cO^\perp$ does not contain any product operator.
	\end{definition}
	
	One can see that the strongly UPUOB is a UPUOB, while the converse fails. We refer authors to FIG. \ref{fig:isomorphism} for the connection among UPOBs, UPUOBs and strongly UPUOBs. Specifically, we give an example of UPUOB on $\bbC^2 \otimes \bbC^3$ which is not a strongly UPUOB in Example \ref{ex:UPUOB23}.
	
	When $n=1$, the UPUOB reduces to the unextendible unitary operator (UUO) set $\{U_i\}_{i=1}^N$ on $\bbC^d$ \cite{PhysRevA.84.042306}. The conditions of a UUO set can be expressed as (i) $U_i \in \mathbb{M}_{d,d}$'s are unitary; (ii) $\tr (U_i^\dagger U_j)=d\delta_{ij}, i,j=1,\cdots, N$; and (iii) if $\tr(U_i^\dagger U)=0$ for all $i=1,\cdots,N$, then $U$ is not a unitary matrix. The UUO set is equivalent to the unextendible maximally entangled bases (UMEBs) \cite{PhysRevA.84.042306}. It is a set of orthonormal maximally entangled states $\{\ket{\psi_i}=\frac{1}{\sqrt{d}}(I\otimes U_i)\sum_{j=1}^d\ket{j,j} | i=1,\cdots,N\}$ in $\bbC^d \otimes \bbC^{d}$ that consists of $N<d^2$ vectors which have no additional maximally entangled state orthogonal to all of them. The following lemma has been investigated in Theorem 1 of \cite{PhysRevA.90.034301,PhysRevA.94.052302}. We restate it using UUO sets.
	
	\begin{lemma}
		\label{le:UUO_qd}
		If there is a UUO set of cardinality $N$ on the space $\bbC^d$, then for any $q=1,\cdots,n$, there exists a UUO set of cardinality $q^2d^2-qd^2+qN$ on the space $\bbC^{qd}$.
	\end{lemma}
	
	This lemma gives a method to construct high dimensional UUO set from lower dimensional UUO set. Furthermore, we can obtain the UPUOB by using the tensor product of the UUO sets and the complete unitary operator sets. In the following, we propose some properties of the tensor product of UPUOBs. 
	
	\begin{lemma}
		\label{le:UPUOagain}
		(i)	The tensor product of UUO sets is always a UPUOB. 
		
		(ii) The tensor product of UPUOBs is still a UPUOB.
		
		(iii) The tensor product of strongly UPUOBs is still a strongly UPUOB.	
	\end{lemma}
	
	\begin{proof}	
		(i) We prove it by contradiction. Let $\{U_i\}_{i=1}^{n_1}, \{U_j'\}_{j=1}^{n_2}$ be UUO sets on the spaces $\bbC^{d_1}$ and $\bbC^{d_2}$, respectively, where $n_1\leq d_1^2,n_2\leq d_2^2$. Suppose there is a product unitary operator $U_{n_1+1}\otimes U'_{n_2+1} \in \mathbb{M}_{d_1,d_1}\otimes \mathbb{M}_{d_2,d_2}$ orthogonal to the set $\{U_i\otimes U_j'\}_{i,j=1}^{n_1,n_2}$. That is, $\tr((U_{n_1+1}\otimes U_{n_2+1}')^\dagger (U_i\otimes U_j'))=0$. So we have $\tr(U_{n_1+1}^\dagger U_i) \tr ((U_{n_2+1}')^\dagger U_j')=0$ for any $i,j$. It is a contradiction with the fact that $\{U_i\}_{i=1}^{n_1}, \{U_j'\}_{j=1}^{n_2}$ are UUO sets on the space $\bbC^{d_1}$ and $\bbC^{d_2}$, respectively.
		
		The above proof can be straightforwardly extended to multipartite UPUOBs. We have proven assertion (i).
		
		(ii) It suffices to prove that the tensor product of two bipartite UPUOBs is still a UPUOB. Let $\{A^{(1)}_{i_1}\otimes A^{(2)}_{i_2}\}_{i_1,i_2=1}^{n_1,n_2}$ and $\{B^{(1)}_{j_1}\otimes B^{(2)}_{j_2}\}_{j_1,j_2=1}^{m_1,m_2}$ be two bipartite UPUOBs on the space $\bbC^{d_1} \otimes \bbC^{d_2}$ and $\bbC^{d_3} \otimes \bbC^{d_4}$, respectively. Next, we prove by contradiction that $\{A^{(1)}_{i_1}\otimes A^{(2)}_{i_2} \otimes B^{(1)}_{j_1}\otimes B^{(2)}_{j_2}\}$ is still a UPUOB on $\bigotimes_{k=1}^4 \bbC^{d_k}$. Suppose there exists a product unitary operator $A^{(1)}_{n_1+1}\otimes A^{(2)}_{n_2+1} \otimes B^{(1)}_{m_1+1} \otimes B^{(2)}_{m_2+1} \in \bigotimes_{k=1}^4\mathbb{M}_{d_k,d_k}$ orthogonal to the set $\{A^{(1)}_{i_1}\otimes A^{(2)}_{i_2} \otimes B^{(1)}_{j_1}\otimes B^{(2)}_{j_2}\} $. Since $\{A^{(1)}_{i_1}\otimes A^{(2)}_{i_2}\}_{i_1,i_2=1}^{n_1,n_2}$ is a bipartite UPUOB, we have $\tr ((A^{(1)}_{n_1+1}\otimes A^{(2)}_{n_2+1})^\dagger (A^{(1)}_{i_1}\otimes A^{(2)}_{i_2})) \neq 0$. Then $B^{(1)}_{m_1+1} \otimes B^{(2)}_{m_2+1}$ must be orthonormal to $ B^{(1)}_{j_1}\otimes B^{(2)}_{j_2}$. It is a contradiction with the fact that $\{B^{(1)}_{j_1}\otimes B^{(2)}_{j_2}\}_{j_1,j_2=1}^{m_1,m_2}$ is a bipartite UPUOB. Thus $\{A^{(1)}_{i_1}\otimes A^{(2)}_{i_2} \otimes B^{(1)}_{j_1}\otimes B^{(2)}_{j_2}\} $ is a UPUOB on the space $\bigotimes_{k=1}^4 \bbC^{d_k}$.

		(iii) One can prove the claim using the idea of (ii). This completes the proof.
	\end{proof}
	
	We can construct UPUOBs based on Lemma \ref{le:UPUOagain} by using UUO sets\cite{PhysRevA.89.062313, PhysRevA.90.034301} and complete unitary operators sets. Using Lemmas \ref{le:UUO_qd} and \ref{le:UPUOagain} (i), one can straightforwardly obtain the following lemma.
	
	\begin{lemma}
		\label{le:UPUOB_qd}
		If there are UUO sets $\mathbb{U}_i$ of cardinality $N_i$ on the space $\bbC^{d_i},i=1,\cdots,k$, then for any $q_i=1,\cdots,n_i$, there exists a UPUOB of cardinality $\prod_{i=1}^k (q_i^2d_i^2-q_id_i^2+q_iN_i)$ which is the tensor product of $\mathbb{U}_i$ on the space $\bigotimes_{i=1}^k\bbC^{q_id_i}$.
	\end{lemma}

	Next, we prove that arbitrary tensor products of bipartite UPOBs are still bipartite UPOBs in Theorem \ref{thm:UPUOBtensor}. Before Theorem \ref{thm:UPUOBtensor}, we propose a preliminary fact 
	on local dimensions extended from Lemma 1 in \cite{PhysRevLett.82.5385}. The proof of the fact uses the idea of the same lemma, and we present it here for completeness.  
	
	\begin{lemma}
		\label{le:local_rank}
		Let $\mathbb{S}=\{(A_j=\bigotimes_{i=1}^m A_{i,j}):j=1,\cdots,n\}$ be an orthogonal product operator basis spanning a subspace of $\bigotimes_{i=1}^m \mathbb{O}(\cH_i)$ with $\dim \cH_{i}=d_i$. Assume that $P$ is a partition of $\mathbb{S}$ into $m$ disjoint subsets: $\mathbb{S}=\mathbb{S}_1 \cup \mathbb{S}_2  \cdots \cup \mathbb{S}_m$. Let $\mathcal{A}_i=\lin \{ A_{i,j}: A_j \in \mathbb{S}_i\}$ and $r_i=\dim \mathcal{A}_i$ be the local dimension of subset $\mathbb{S}_i$ of the $i$-th party. Then $\mathbb{S}$ is extendible if and only if there exists a partition $P$ such that for all $i=1,\cdots,m$, the local dimension $r_i$ of the $i$-th subset is less than $d_i$.	
	\end{lemma}
	\begin{proof}
		By definition, $\mathbb{S}$ is extendible if and only if there exists a product operator $W$ orthogonal to $\bbS$. Equivalently, a partition can be found such that $W$ is orthogonal to all the operators in $\mathbb{S}_1$ for party $1$, all the operators in $\mathbb{S}_2$ for party $2$, and so on through $\mathbb{S}_m$. This can be done if each of the sets $\mathbb{S}_i$ has local dimension $r_i$ less than the dimension $d_i$ of the $i$-th party's Hilbert space. Conversely, if there exists at least one of the sets $\mathbb{S}_i$ having full local dimension $r_i$ equal to $d_i$, then there is no way to choose $W$ orthogonal to $\bbS$. Thus the original set $\mathbb{S}$ is unextendible. This completes the proof. 
	\end{proof}
	
	If the $A_j$'s are vectors then one can see that Lemma \ref{le:local_rank} reduces to Lemma 1 in \cite{PhysRevLett.82.5385}. Due to the bijection $\ketbra{i}{j} \lra \ket{i,j}$ between matrices and vectors, every UPOB in $\bbM_{a,b}\otimes\bbM_{c,d}$ one-to-one corresponds to a bipartite UPB in $\bbC^{ab}\otimes\bbC^{cd}$. Hence, the following observation can be obtained from Theorem 8 of \cite{ISI:000184088700001} on the version of UPBs, though a strightforward proof would require the use of Lemma \ref{le:local_rank}.
	
	\begin{theorem}
		\label{thm:UPUOBtensor}
		Let two bipartite UPOBs be $S_1=\{A_i\}_{i=1}^{l_1}$ on $\bbC^{d_1} \otimes \bbC^{d_2}$, and $S_2=\{A_i'\}_{i=1}^{l_2}$ on $\bbC^{d_3} \otimes \bbC^{d_4}$. The tensor product operator basis $\{A_i\otimes A_j'\}_{i,j=1}^{l_1,l_2}$ is a bipartite UPOB on $\bbC^{d_1d_3}\otimes \bbC^{d_2d_4}$.
	\end{theorem}
	
	This theorem has the consequence that arbitrary tensor powers of bipartite UPOBs are again UPOBs. Further, a generalization of this theorem for multipartite UPOBs still holds.

	\section{Construction of UPOBs, UPUOBs and strongly UPUOBs}
	\label{sec:construction}
	
	In this section, we construct UPOBs and strongly UPUOBs. A bijection from vectors to matrices is proposed to construct UPOBs from UPBs. We give an example of UPOB on $\bbC^2 \otimes \bbC^2$ constructed from a $4 \times 4$ UPB using this bijection in Example \ref{ex:UPB-UPOB}. Further, Proposition \ref{pro:UPOB_size} shows the existence of some size of UPOBs. We also show that there exists a strongly UPUOB on $(\bbC^2)^{\otimes n}$ for $n \geq 2$ in Theorem \ref{thm:UPUOB2n}. In Theorem \ref{thm:UPUOB22d_i}, we give a construction of $(n+2)$-partite UPUOB on the space $(\bbC^2)^{\otimes 2} \otimes (\bigotimes_{i=1}^n \bbC^{d_i})$ with $n\geq 0$. In Example \ref{ex:UPUOB23} and Theorem \ref{thm:UPUOBqd}, we show that a set of bipartite UPUOB is not strongly UPUOBs.

	To begin with, we define a bijection $\cF$ from the set of $d$-dimensional vectors $\ket{a^{(j)}}:=(a^{(j)}_1,a^{(j)}_2,\cdots,a^{(j)}_d)^T $ to the matrices $[c_{p,q}^{(j)}] \in \mathbb{M}_{m,n}$, $j=1,\cdots,N$ as follows,
	\begin{eqnarray}
		\label{eq:bijection_general}
		\cF: \ket{a^{(j)}}\mapsto [c_{p,q}^{(j)}].
	\end{eqnarray}
	We choose the set 
	\begin{eqnarray}
		\label{eq:setA}
		\mathcal{A}=\{f_1,f_2,\cdots,f_{d}\} \subseteq \mathcal{S}=\{(p,q), \\ \notag
		p=1,\cdots,m,q=1,\cdots,n\}.
	\end{eqnarray}
	Let $a^{(j)}_1=c_{f_1}^{(j)},a^{(j)}_2=c_{f_2}^{(j)},\cdots,a^{(j)}_d=c_{f_{d}}^{(j)}$, and other $c_{p,q}^{(j)}$ be zero entries, where $\max\{m,n\}\leq d \leq m n$. With the bijection $\cF$ in \eqref{eq:bijection_general}, we can obtain the matrices $M^{(1)},M^{(2)} \in \mathbb{M}_{m,n}$ from the $d$-dimensional vectors $\ket{a^{(1)}}, \ket{a^{(2)}}$, respectively. One can verify that $M^{(1)},M^{(2)}$ are orthonormal if and only if $\ket{a^{(1)}}, \ket{a^{(2)}}$ are orthonormal. 
	
	One can construct a UPOB $\{\bigotimes_{i=1}^k M^{(j)}_{m_i,n_i}\}$ from a $k$-partite UPB $\{\bigotimes_{i=1}^k \ket{a_{i}^{(j)}}\}$ in $\bigotimes_{i=1}^k \bbC^{d_i}$ with cardinality $N$ using the bijection $\cF$ for each party, where $j=1,\cdots,N$ and $d_i=m_i n_i$ for $m_i,n_i>1$. For the $k$-partite UPB $\{\bigotimes_{i=1}^k \ket{a_{i}^{(j)}}\}$ in $\bigotimes_{i=1}^k \bbC^{d_i}$, it has been proved that $N \leq \prod_{i=1}^k d_i-4$ in \cite{Chen_2013}. Thus, we also know the bound of the cardinality of multipartite UPOB is $\prod_{i=1}^k d_i-4$ using this bijection $\cF$. There are also some results on the bound of bipartite UPB in terms of Ramsey number. For example, the cardinality of a UPB in $\bbC^3 \otimes \bbC^3$ is smaller than the Ramsey number $R(3,3)=6$ \cite{ISI:000184088700001}. However, the relationship between the cardinality of multipartite UPOB and Ramsey number remains unknown. Further, we say a UPB $\{\bigotimes_{i=1}^k \ket{a^{(j)}_i}\}$ in $\bigotimes_{i=1}^k \bbC^{d_i}$ is a strongly UPB if $\sum_{x=0}^{d_i-1}\ket{a^{(j)}_i}_{d_ix+t}\overline{\bra{a^{(j)}_i}}_{d_ix+s}=\delta_{s,t}$, where $\delta_{s,t}=1$ if $s=t$ and $0$ otherwise, $s,t=0,1,\cdots,d_i-1$. The strongly UPUOB can be obtained from the strongly UPB using the bijection $\cF$ for each party.
	
	In the following, We give an example of UPOB on $\mathbb{C}^2 \otimes \mathbb{C}^2$ from a $4\times4$ UPB.
	
	\begin{example}
		\label{ex:UPB-UPOB}
		Consider the UPB in Example \ref{ex:UPB}. Since $d_i=4=2\times 2$, we can construct UPOB on $\mathbb{C}^2 \otimes \mathbb{C}^2$ using the bijection $\cF$. For both parties, using the set $\mathcal{A}=\{f_1=(1,1),f_2=(1,2),f_3=(2,1), f_4=(2,2)\}$ in \eqref{eq:setA}, we can obtain the UPOB $\{M_j,j=1,\cdots,11\}$ on $\mathbb{C}^2 \otimes \mathbb{C}^2$, where
		\begin{eqnarray}
			\notag
			\begin{aligned}
				& M_1=\frac{1}{\sqrt{2}}\bma
				1 & 0\\
				0 & 0
				\ema \otimes
				\bma
				1& -1\\
				0& 0 
				\ema, 
				&& M_2=\frac{1}{\sqrt{2}}\bma
				1 & 0\\
				0 & -1
				\ema \otimes
				\bma
				0& 0\\
				1& 0 
				\ema, \\
				& M_3=\frac{1}{\sqrt{3}}\bma
				1 & \omega_3\\
				\omega_3^2 & 0
				\ema \otimes
				\bma
				0& 0\\
				0& 1 
				\ema, 
				&& M_4=\frac{1}{\sqrt{3}}\bma
				1 & \omega_3\\
				\omega_3^2 & 0
				\ema \otimes
				\bma
				0& 0\\
				0& 1 
				\ema, \\
				& M_5=\frac{1}{\sqrt{3}}\bma
				0 & 1\\
				\omega_3 & \omega_3^2
				\ema \otimes
				\bma
				0& 1\\
				0& 0 
				\ema, 
				&& M_6=\frac{1}{\sqrt{3}}\bma
				0 & 1\\
				\omega_3^2 & \omega_3
				\ema \otimes
				\bma
				0& 1\\
				0& 0 
				\ema, \\
				& M_7=\frac{1}{\sqrt{2}}\bma
				0 & 0\\
				0 & 1
				\ema \otimes
				\bma
				1& 0\\
				0& -1 
				\ema, 
				&& M_8=\frac{1}{2}\bma
				0 & 1\\
				1 & 0
				\ema \otimes
				\bma
				1& 0\\
				-1& 0 
				\ema, \\
				& M_9=\frac{1}{2}\bma
				0 & 1\\
				-1 & 0
				\ema \otimes
				\bma
				1& 0\\
				1& 0 
				\ema, 
				&& M_{10}=\frac{1}{2}\bma
				0 & 1\\
				-1 & 0
				\ema \otimes
				\bma
				1& 0\\
				-1& 0 
				\ema, \\
				&M_{11}=\frac{1}{4}\bma
				1& 1 \\
				1& 1 
				\ema \otimes 
				\bma
				1 & 1\\
				1 & 1
				\ema.
			\end{aligned}
		\end{eqnarray}
	\end{example}
	
	Using the above idea, we extend Example \ref{ex:UPB-UPOB} to the UPOB on $\mathbb{C}^m \otimes \mathbb{C}^n$ from a UPB in $\mathbb{C}^{m^2}\otimes \mathbb{C}^{n^2}$. 
	
	\begin{proposition}
		\label{pro:UPOB_size}
		(i) There exists a UPOB of size $d_1d_2-4\lfloor \frac{d_1-1}{2}\rfloor$ in $\mathbb{M}_{m_1,n_1} \otimes \mathbb{M}_{m_2,n_2}$, where $3\leq d_1 \leq d_2$ and $d_1=m_1n_1,d_2=m_2n_2$.
		
		(ii) There exists a UPOB of size $d_1d_2-k$ in $\mathbb{M}_{m_1,n_1} \otimes \mathbb{M}_{m_2,n_2}$ for $4 \leq k \leq 2d_1-1$, where $4\leq d_1 \leq d_2$ and $d_1=m_1n_1,d_2=m_2n_2$.
	\end{proposition}
	
	\begin{proof}
		(i) Since there exists a UPB of size $d_1d_2-4\lfloor \frac{d_1-1}{2}\rfloor$ in $\mathbb{C}^{d_1} \otimes \mathbb{C}^{d_2}$ for $3\leq d_1 \leq d_2$ \cite{PhysRevA.101.062329}, one can obtain a UPOB in $\mathbb{M}_{m_1,n_1} \otimes \mathbb{M}_{m_2,n_2}$ of size $d_1d_2-4\lfloor \frac{d_1-1}{2}\rfloor$ on $\mathbb{C}^{d_1} \otimes \mathbb{C}^{d_2}$ by using the bijection $\cF$ in \eqref{eq:bijection_general} in every partition, where $d_1=m_1n_1,d_2=m_2n_2$. 
		
		(ii) It has been proved that there exists a UPB of size $d_1d_2-k$ in $\mathbb{C}^{d_1} \otimes \mathbb{C}^{d_2}$ for $4 \leq k \leq 2d_1-1$ and $4 \leq d_1 \leq d_2$ \cite{PhysRevA.101.062329}. Thus one can construct a UPOB in $\mathbb{M}_{m_1,n_1} \otimes \mathbb{M}_{m_2,n_2}$ of size $d_1d_2-4\lfloor \frac{d_1-1}{2}\rfloor$ on $\mathbb{C}^{d_1} \otimes \mathbb{C}^{d_2}$ by using the bijection $\cF$ in \eqref{eq:bijection_general} in every partition, where $d_1=m_1n_1,d_2=m_2n_2$. This completes the proof.	
	\end{proof}
	
	Since there do not exist qubit UUO sets \cite{PhysRevA.84.042306}, we cannot obtain UPUOB on the space $(\bbC^2)^{\otimes n}$ using tensor products of qubit UUO sets according to Lemma \ref{le:UPUOagain}. In the following, we give another method to construct strongly UPUOBs on the space $(\bbC^2)^{\otimes n}$.
	
	\begin{theorem}
		\label{thm:UPUOB2n}
		There exists a strongly UPUOB on the space $(\bbC^2)^{\otimes n}$ for any $n \geq 2$.
	\end{theorem}
	\begin{proof}
		It has been proved that $\mathbb{U}_2:=\{U_i\}_{i=1}^{12}$ is a UUO set on the space $\bbC^4$ \cite{PhysRevA.84.042306}, where
		\begin{eqnarray}
			\label{eq:U1-5}
			&&
			U_1=\frac{1}{\sqrt{2}}\sigma_x \otimes(\sigma_x-\sigma_y), \notag \\
			&&
			U_2=\frac{1}{\sqrt{2}}(\sigma_x-\sigma_y) \otimes \sigma_z,\notag	\\
			&&
			U_3=\frac{1}{\sqrt{2}} \sigma_z\otimes (-\sigma_y+\sigma_z), \notag \\
			&&
			U_4=\frac{1}{\sqrt{2}}(-\sigma_y+\sigma_z) \otimes \sigma_x, \notag	\\
			&&
			U_5=\frac{1}{3}(\sigma_x+\sigma_y+\sigma_z) \otimes (\sigma_x+\sigma_y+\sigma_z),
		\end{eqnarray}
		and 
		\begin{eqnarray}
			\label{eq:U6-12}
			&&
			U_6, \cdots,U_{12}=I\otimes I, I \otimes \sigma_{\alpha}, \sigma_{\beta}\otimes I, 
		\end{eqnarray}
		and $\alpha,\beta=x,y,z$. We prove that $\mathbb{U}_2$ is a strongly UPUOB of systems $A, B$ on the space $\bbC^2 \otimes \bbC^2$. Assume that there exists a two-qubit product operator $W \otimes X$ on the space $\bbC^2 \otimes \bbC^2$ which is orthonormal to the set $\mathbb{U}_2$. Since $W \otimes X$ is orthonormal to the operators in \eqref{eq:U6-12}, we obtain that $W,X$ must be the linear combination of $\sigma_\alpha$ and $\a=x,y,z$. Further, because $W \otimes X$ is orthonormal to the operators in \eqref{eq:U1-5}, one can obtain that $W \otimes X $ is $k$-orthonormal to three of $U_1,\cdots,U_5$ in \eqref{eq:U1-5}, $k=1$ or $2$. Note that the $A$-partition of any three of $U_1,\cdots,U_5$ are linearly independent and they are the linear combination of $\sigma_\alpha, \a=x,y,z$, and the $B$-partition of $U_1,\cdots,U_5$ has the same property. So we have $W=0$ or $X=0$. It is a contradiction with the definition of strongly UPUOBs. Thus $\mathbb{U}_2$ is a two-qubit strongly UPUOB. 
		
		Next we construct a strongly UPUOB on the space $(\bbC^2)^{\otimes n}$ by using $\mathbb{U}_2$. Let $\mathbb{S}=\{ \sigma_{\a_1}\otimes \cdots \otimes \sigma_{\a_{n-2}}\otimes U_i\}$, where $\a_1,\cdots, \a_{n-2}=0,x,y,z$, the matrix $\sigma_0=I$ and $U_i, i=1,\cdots,12$ are UUOs in $\mathbb{U}_2$. Using Lemma \ref{le:UPUOagain} (iii), one can verify that $\mathbb{S}$ is a strongly UPUOB with $|\mathbb{S}|=3\cdot4^{n-1}$ on the space $(\bbC^2)^{\otimes n}$. We have proven the assertion.
	\end{proof}
	
	Similarly to the last paragraph of the proof of Theorem \ref{thm:UPUOB2n}, one can construct a UPUOB on the space $(\bbC^2)^{\otimes 2} \otimes (\bigotimes_{i=1}^n \bbC^{d_i})$.
	
	\begin{theorem}
		\label{thm:UPUOB22d_i}
		There exists an $(n+2)$-partite UPUOB on the space $(\bbC^2)^{\otimes 2} \otimes (\bigotimes_{i=1}^n \bbC^{d_i})$, $n\geq 0$.
	\end{theorem} 
	\begin{proof}
		If there do not exist a UUO set on $\bbC^{d_i}$, then we can choose a complete operator basis for the $i+2$ parties. Using the strongly UPUOB $\mathbb{U}_2$ in \eqref{eq:U1-5}-\eqref{eq:U6-12}, Lemmas \ref{le:UUO_qd}-\ref{le:UPUOB_qd}, one can prove this assertion.
	\end{proof}
	
	The strongly UPUOBs must be UPUOBs on the space $(\bbC^2)^{\otimes n}$ according to the definitions. In the following, we give an example of UPUOB on $\bbC^2 \otimes \bbC^3$ which is not a strongly UPUOB.
	
	\begin{example}
		\label{ex:UPUOB23}
		Let
		\begin{eqnarray}
			\label{eq:UUO_6}
			&&
			U_{n,m}^{\pm}=\xi_{\pm}\otimes U_{n,m}=\xi_{\pm}\otimes \sum_{k=0}^2(\omega_3)^{kn}\ketbra{k \oplus m}{k}, n,m=1,2,3, \notag \\
			&&
			U_s^{\pm}=\eta_{\pm} \otimes U_s=\eta_{\pm} \otimes (I-(1-e^{i \theta})\ketbra{\psi_s}{\psi_s}), s=1,\cdots,6, \notag \\
		\end{eqnarray}
		where $\omega_3=e^{\frac{2\p i}{3}}$, $\xi_{\pm}=\bma
		0 & 1 \\
		\pm1 & 0
		\ema$, $\eta_{\pm}=\bma
		1 & 0\\
		0 & \pm1
		\ema$, $k\oplus m$ denotes $(k+m)$ mod $3$, $\cos \theta =-\frac{7}{8}$ and
		\begin{eqnarray}
			&&
			\ket{\psi_{1,2}}=\frac{1}{\sqrt{1+\phi^2}}(\ket{0} \pm \phi \ket{1}), \notag \\
			&&
			\ket{\psi_{3,4}}=\frac{1}{\sqrt{1+\phi^2}}(\ket{1} \pm \phi \ket{2}),\notag \\
			&&
			\ket{\psi_{5,6}}=\frac{1}{\sqrt{1+\phi^2}}(\ket{2} \pm \phi \ket{0}), \notag
		\end{eqnarray} 
		$\phi=\frac{1+\sqrt{5}}{2}$. This set $\{U_{n,m}^{\pm},U_s^{\pm}\}_{n,m,s=1}^{3,3,6}$ is a UUO set on the space $\bbC^6$ \cite{PhysRevA.90.034301}. It is also a UPUOB on the space $\bbC^2 \otimes \bbC^3$, because there do not exist a unitary operator in its complementary space orthogonal to the set. However, it is not a strongly UPUOB, because of the claim that one can find a bipartite product operator $\tilde{U}=W \otimes X$ on the space $\bbC^2 \otimes \bbC^3$ orthogonal to the set $\{U_{n,m}^{\pm},U_s^{\pm}\}_{n,m,s=1}^{3,3,6}$. 	The proof of this claim is as below. 
		
		\begin{proof}
			Suppose that the operator $\tilde{U}=W \otimes X$ satisfies $\tr((\tilde{U})^\dagger U_{n,m}^{\pm})=0$ and $\tr((\tilde{U})^\dagger U_s^{\pm})=0$. Let the three matrix subspaces $\mathcal{V}_1=\mathrm{span} \{U_{n,m}^{\pm}\}$, $\mathcal{V}_2=\{\bma
			A & \mathbf{0}\\
			\mathbf{0} & B
			\ema|A,B \in \mathbb{M}_{3,3}  \}$ and $\mathcal{V}_3=\mathrm{span}\{U_{n,m}^\pm, U_s^\pm\}$ in \eqref{eq:UUO_6}. Then we have $\dim (\mathcal{V}_1)=\dim (\mathcal{V}_2)=18$, and $\dim(\mathcal{V}_3)=30$. Since $\tr (\bma
			A & \mathbf{0}\\
			\mathbf{0} & B
			\ema^\dagger 
			\bma
			\mathbf{0} & U_{n,m} \\
			\pm U_{n,m} & \mathbf{0}
			\ema)=0$, we can obtain $\mathcal{V}_1^\perp=\mathcal{V}_2$, where $\mathcal{V}_1^\perp$ is the complementary space of $\mathcal{V}_1$. So we have $\mathcal{V}_3^\perp \subset \mathcal{V}_1^\perp=\mathcal{V}_2$. Thus $\tilde{U}=W \otimes X \in \mathcal{V}_3^\perp=\mathcal{V}_2$ and must be the form $\tilde{U}=W \otimes X=\bma
			w_1 X & w_2 X\\
			w_3X & w_4X
			\ema=\bma
			w_1 X &0\\
			0& w_4X
			\ema$. As $\tilde{U}=W \otimes X$ is orthogonal to $U_s^\pm$ in \eqref{eq:UUO_6}, we have $
			\tr(\bma
			w_1 X &0\\
			0& w_4X
			\ema^\dagger 
			\bma
			U_s & \mathbf{0} \\
			\mathbf{0}& \pm U_s
			\ema)=0$. We can obtain that $\tr(w_1^\ast X^\dagger U_s) \pm \tr(w_4^\ast X^\dagger U_i)=0$. So $\tr(w_1^\ast X^\dagger U_s)= \tr(w_4^\ast X^\dagger U_s)=0$. Since $U_s$ in \eqref{eq:UUO_6} are symmetric matrices in the computational basis $\ket{0}, \ket{1}, \ket{2}$, we obtain that $w_1X$ and $w_4X$ are in the antisymmetric subspace represented by the computational basis $\ket{0}, \ket{1}, \ket{2}$. Thus $(w_1X)^T=-w_1X, (w_4X)^T=-w_4X$. As $w_1 w_4 \neq 0$, we have $X^T=-X$. That is, $X$ is an anti-symmetric matrix. So we have constructed a bipartite product operator $\tilde{U}=W \otimes X$ orthogonal to the set $\{U_{n,m}^{\pm},U_s^{\pm}\}_{n,m,s=1}^{3,3,6}$. This completes the proof.
			
		\end{proof}
		
	\end{example}
	
	In the following, we re-derive the construction by using the idea in Example \ref{ex:UPUOB23}. It has been proved that one can construct a UUO set on $\bbC^{qd}$ from the UUO set on $\bbC^d$ for any $q=1,2,\cdots$ \cite{PhysRevA.90.034301}. Let the order-$q$ matrix 
	\begin{eqnarray}
		P_q=\bma
		0 & 1 & 0 & \cdots & 0 \\ 
		0 & 0 & 1 & \cdots & 0\\
		\vdots & \vdots & \vdots & \ddots &\vdots \\ 
		0 & 0 & 0 & \cdots & 1\\
		1 & 0 & 0 & \cdots & 0
		\ema, W_q=\bma
		1 & 1 & 1 & \cdots & 1\\
		1 & \omega_q & \omega_q^2 & \cdots & \omega_q^{q-1} \\
		1 & \omega_q^2 & \omega_q^4 & \cdots & \omega_q^{2(q-1)}\\
		\vdots & \vdots & \vdots & \ddots & \vdots \\
		1 & \omega_q^{q-1} & \omega_q^{2(q-1)} & \cdots & \omega_q^{(q-1)^2}
		\ema, \notag
	\end{eqnarray}
	where $\omega_q=e^{\frac{2\pi i}{q}}$. We denote $U_{n,m}=\sum_{k=0}^{d-1}(\omega_d)^{kn}\ketbra{k \oplus m}{k}, m,n=0,1,\cdots,d-1$, and $k \oplus m= (k+m) \mod d$. Let $\{U_t\}_{t=1}^{N<d^2}$ be a UUO set on $\bbC^d$. Then we set 
	\begin{eqnarray}
		\label{eq:UPUOB_nmsj}
		&&
		U_{n,m}^{(s,j)}=(W_q^sP_q^j)\otimes U_{n,m},\notag \\
		&&
		U_t^{(s)}=W_q^s \otimes U_t,
	\end{eqnarray} 
	where $s=0,1, \cdots, q-1, j=1,2,\cdots,q-1, m,n=0,1,\cdots,d-1$, and $t=1,2,\cdots,N<d^2$. It has been proven that $\{U_{n,m}^{(s,j)},U_t^{(s)}\}$ is a UUO set on $\bbC^{qd}$ with cardinality $q^2d^2-qd^2+qN$ constructed from the UUO set $\{U_t\}$ on $\bbC^d$ \cite{PhysRevA.90.034301}. Similar to Example \ref{ex:UPUOB23}, we present the following observation.
	
	\begin{theorem}
		\label{thm:UPUOBqd}
		If there exists a UUO set on $\bbC^d$, then the set $\{U_{n,m}^{(s,j)},U_t^{(s)}\}$ in \eqref{eq:UPUOB_nmsj} is a UPUOB but not a strongly UPUOB on $\bbC^q \otimes \bbC^d$.
	\end{theorem}
	\begin{proof}
		Let $\{U_t\}_{t=1}^{N<d^2}$ be the UUO set on $\bbC^d$. Then one can construct the UUO set on the space $\bbC^{qd}$ \cite{PhysRevA.90.034301} as shown in \eqref{eq:UPUOB_nmsj}. This UUO set $\{U_{n,m}^{(s,j)},U_t^{(s)}\}$ is a UPUOB on the space $\bbC^q \otimes \bbC^d$, because there do not exist a unitary operator in its complementary space which is orthogonal to the set \cite{PhysRevA.90.034301}. However, it is not a strongly UPUOB, because there exists a bipartite product operator $\tilde{U'}=W_q' \otimes X'_d$ on the space $\bbC^q \otimes \bbC^d$ which is orthogonal to the set $\{U_{n,m}^{(s,j)},U_t^{(s)}\}$. The proof is similar to the claim of Example \ref{ex:UPUOB23}. In particular, $W_q'$ must be the form $W_q'=\diag(w_1',\cdots,w_q')$ and $X'_d$ must be an anti-symmetric matrix.
	\end{proof}
	
	Thus from the above Example \ref{ex:UPUOB23} and Theorem \ref{thm:UPUOBqd}, we can see the connection between UPUOBs and strongly UPUOBs. Similarly to the application of UPBs, we ask whether we can distinguish UPOBs or strongly UPUOBs by LOCC. This will be discussed in the next section.

	\section{Application: Nonlocality of UPUOB}
	\label{sec:app}
	
\begin{figure}[h] 
		\center{\includegraphics[width=8.5cm]  {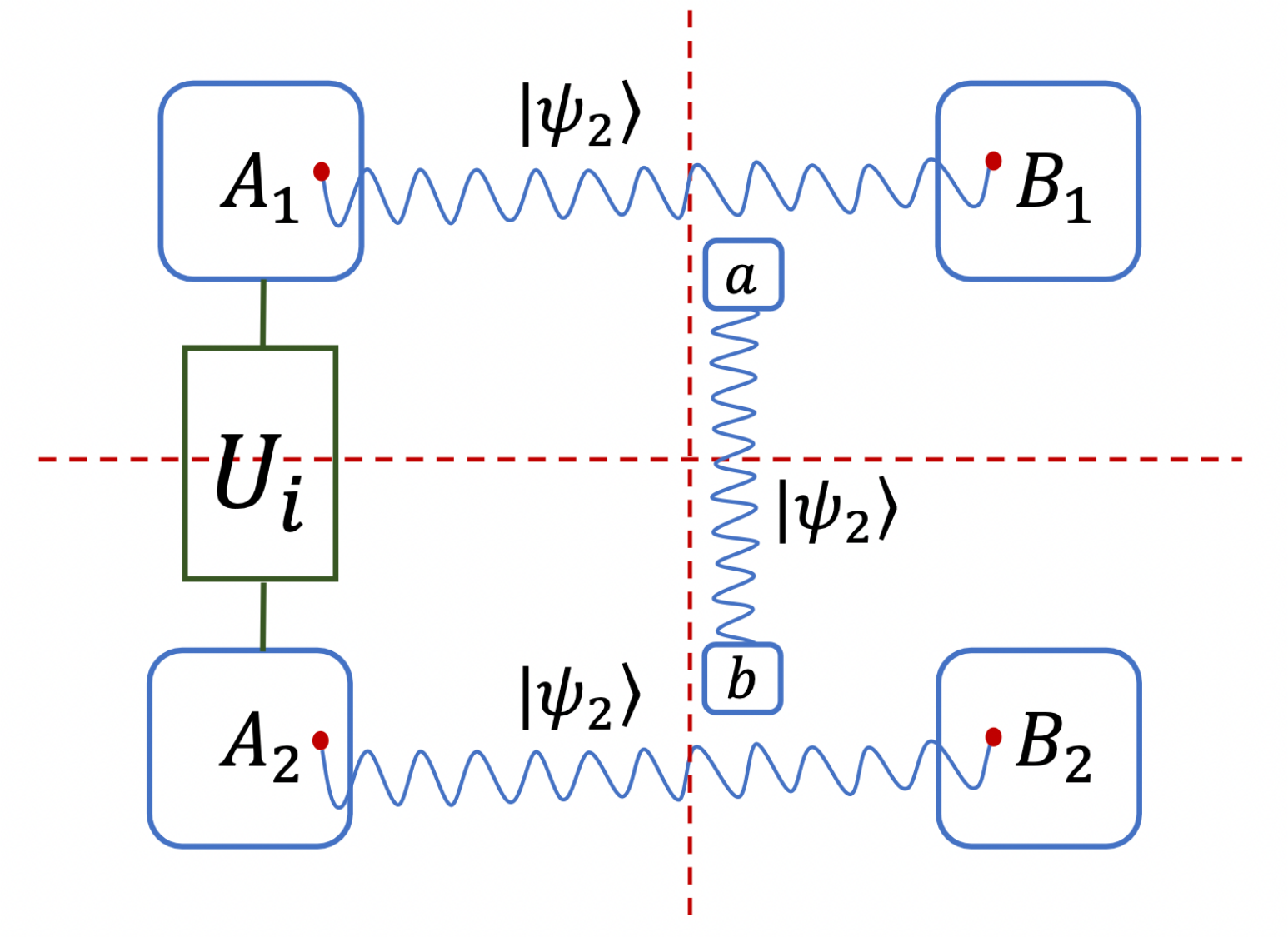}} 
		\caption{\label{fig:discriminationU2} $A_1$ and $B_1$ share a two-level MES $\ket{\psi_2}$, and $A_2$ and $B_2$ also share a two-level MES $\ket{\psi_2}$. The operators $\{U_i\}_{i=1}^{12}$ are the strongly UPUOB in \eqref{eq:U1-5}-\eqref{eq:U6-12}. Availability of additional entanglement resources across the vertical and/or the horizontal dotted lines give rise to different sets of allowed operations performed by $A_i$ and $B_i,i=1,2$. When we perform the operation $U_i$ on $A_1$ and $A_2$, then as shown in Theorem \ref{thm:ebitent}, it only requires a two-level MES $\ket{\psi_2}$ between the ancillary systems $a$ and $b$ to distinguish $U_i,i=1,\cdots,12$. Thus it requires three ebits of entanglement resources to distinguish $U_i,i=1,\cdots,12$.} 
	\end{figure}

	In this section we prove that the strongly UPUOB $\mathbb{U}_2$ in \eqref{eq:U1-5}-\eqref{eq:U6-12} is genuinely nonlocal in terms of the two-level maximally entangled state (MES). This is presented in Theorem \ref{thm:UPUOB22nonlocal}. Further, Theorem \ref{thm:ebitent} shows that the strongly UPUOB $\mathbb{U}_2$ in \eqref{eq:U1-5}-\eqref{eq:U6-12} can be locally distinguished by using three ebits of entanglement.  As shown in FIG. \ref{fig:discriminationU2}, it has an important application in multipartite secret sharing and the distinguishability of product unitary gates.
	
	Let the $d$-level MES $\ket{\psi_d}=\frac{1}{\sqrt{d}} \sum_{i=1}^d\ket{j,j}$. We define $\ket{a'_k}$ as follows,
	\begin{eqnarray}
		\label{eq:a_kspecial}
		\ket{a'_k}=U_k\otimes I_{B_1,\cdots,B_n}(\ket{\psi_{d}}_{A_1B_1} \otimes \cdots \otimes \ket{\psi_{d}}_{A_nB_n}),	
	\end{eqnarray}
	where $U_k$ are $n$-partite UPUOs of system $A_1,\cdots,A_n$ in $\bigotimes_{i=1}^n \mathbb{C}^{d},k=1,\cdots,N$, and $N$ is the cardinality of the UPUOB $\{U_k\}$. One can verify that the states $\ket{a'_k}$'s are $n$-partite OP states. Next, we show that these states are indistinguishable under LOCC, when $U_k$ in \eqref{eq:a_kspecial} are the genuinely UPUO in \eqref{eq:U1-5}-\eqref{eq:U6-12}. 
	
	\begin{theorem}
		\label{thm:UPUOB22nonlocal}
		The strongly UPUOB $\mathbb{U}_2$ in \eqref{eq:U1-5}-\eqref{eq:U6-12} is genuinely nonlocal in terms of the two-level MES.
	\end{theorem}
	
	\begin{proof}
		Let $U_k$ in \eqref{eq:a_kspecial} be the genuinely UPUO in \eqref{eq:U1-5} and \eqref{eq:U6-12}. According to Definition \ref{def:nonlocaloperator} (i), to show that the strongly UPUOB $\mathbb{U}_2$ is genuinely nonlocal in terms of the two-level MES $\ket{\psi_2}=\frac{1}{\sqrt{2}}(\ket{00}+\ket{11})$, we need to prove that the bipartite state $\ket{a'_k}$ in \eqref{eq:a_kspecial} are nonlocal in terms of the MES. Further, it is impossible to distinguish more than $d'$ MESs in $\bbC^d \otimes \bbC^{d'}$ by LOCC \cite{PhysRevLett.96.040501}. Note that except for the  $A_1B_1|A_2B_2$ bipartition, other bipartitions are all MESs in $\bbC^2 \otimes \bbC^8$ such as $A_1|B_1A_2B_2$ partition, or MESs in $\bbC^4 \otimes \bbC^4$ such as $A_1A_2|B_1B_2$ partition. Since the number of UPUO in \eqref{eq:U1-5} and \eqref{eq:U6-12} is $12$ which is larger than $8$ and $4$, we only need to consider the $A_1B_1|A_2B_2$ bipartition for nonlocality. Next, we show that $\ket{a'_k}$ are locally indistinguishable in $A_1B_1|A_2B_2$ bipartition, where
		\begin{widetext}
			\begin{eqnarray}
				\label{eq:ak22}
				&&
				\ket{a'_1}=(U_1 \otimes I_4) (\ket{\psi_2}\otimes \ket{\psi_2}) \notag \\
				&&=\frac{1}{2\sqrt{2}}(\sigma_x \otimes(\sigma_x-\sigma_y) \otimes I_4) (\ket{0,0}+\ket{1,1})\otimes (\ket{0,0}+\ket{1,1}), \notag \\
				&&
				\ket{a'_2}=(U_2 \otimes I_4) (\ket{\psi_2}\otimes \ket{\psi_2})\notag \\
				&&=\frac{1}{2\sqrt{2}}((\sigma_x-\sigma_y) \otimes \sigma_z \otimes I_4)(\ket{0,0}+\ket{1,1})\otimes (\ket{0,0}+\ket{1,1}) , \notag \\
				&&
				\ket{a'_3}=(U_3 \otimes I_4) (\ket{\psi_2}\otimes \ket{\psi_2})\notag \\
				&&=\frac{1}{2\sqrt{2}}(\sigma_z\otimes (-\sigma_y+\sigma_z)\otimes I_4) (\ket{0,0}+\ket{1,1})\otimes (\ket{0,0}+\ket{1,1}) , \notag \\
				&&		
				\ket{a'_4}=(U_4 \otimes I_4) (\ket{\psi_2}\otimes \ket{\psi_2})\notag \\
				&&=\frac{1}{2\sqrt{2}}((-\sigma_y+\sigma_z) \otimes \sigma_x\otimes I_4)(\ket{0,0}+\ket{1,1})\otimes (\ket{0,0}+\ket{1,1}) , \notag \\
				&&	
				\ket{a'_5}=(U_5 \otimes I_4)(\ket{\psi_2}\otimes \ket{\psi_2}) \notag \\
				&&=\frac{1}{6}((\sigma_x+\sigma_y+\sigma_z) \otimes (\sigma_x+\sigma_y+\sigma_z) \otimes I_4)(\ket{0,0}+\ket{1,1})\otimes (\ket{0,0}+\ket{1,1}) , \notag \\
				&&
				\ket{a'_6}=(U_6 \otimes I_4) (\ket{\psi_2}\otimes \ket{\psi_2})=\frac{1}{2}(I_4\otimes I_4)(\ket{0,0}+\ket{1,1})\otimes (\ket{0,0}+\ket{1,1})) , \notag \\
				&&
				\ket{a'_7}=(U_7 \otimes I_4) (\ket{\psi_2}\otimes \ket{\psi_2})=\frac{1}{2}(I_2 \otimes\sigma_x \otimes I_4)(\ket{0,0}+\ket{1,1})\otimes (\ket{0,0}+\ket{1,1}), \notag \\
				&&
				\ket{a'_8}=(U_8 \otimes I_4) (\ket{\psi_2}\otimes \ket{\psi_2})=\frac{1}{2}(I_2 \otimes\sigma_y \otimes I_4)(\ket{0,0}+\ket{1,1})\otimes (\ket{0,0}+\ket{1,1}) , \notag \\
				&&
				\ket{a'_9}=(U_9 \otimes I_4)(\ket{\psi_2}\otimes \ket{\psi_2})=\frac{1}{2}(I_2 \otimes\sigma_z \otimes I_4)(\ket{0,0}+\ket{1,1})\otimes (\ket{0,0}+\ket{1,1}), \notag \\
				&&
				\ket{a'_{10}}=(U_{10} \otimes I_4) (\ket{\psi_2}\otimes \ket{\psi_2})=\frac{1}{2}(\sigma_x \otimes I_2 \otimes I_4)(\ket{0,0}+\ket{1,1})\otimes (\ket{0,0}+\ket{1,1}), \notag \\
				&&
				\ket{a'_{11}}=(U_{11} \otimes I_4) (\ket{\psi_2}\otimes \ket{\psi_2})=\frac{1}{2}(\sigma_y \otimes I_2 \otimes I_4)(\ket{0,0}+\ket{1,1})\otimes (\ket{0,0}+\ket{1,1}), \notag \\
				&&
				\ket{a'_{12}}=(U_{12} \otimes I_4) (\ket{\psi_2}\otimes \ket{\psi_2})=\frac{1}{2}(\sigma_z \otimes I_2 \otimes I_4)(\ket{0,0}+\ket{1,1})\otimes (\ket{0,0}+\ket{1,1}), \notag \\
			\end{eqnarray} 
		\end{widetext}
		and $U_1, U_{2}, \cdots, U_{12}$ are the UPUO in \eqref{eq:U1-5} and \eqref{eq:U6-12}.
		
		Define a bijection from the basis $\{\ket{p,q}\}_{p,q=0}^1$ in $\bbC^2 \otimes \bbC^2$ to the basis in $\bbC^4$ as follows: $\ket{0,0} \rightarrow \ket{0}, \quad \ket{0,1} \rightarrow \ket{1}, \quad \ket{1,0} \rightarrow \ket{2}, \quad \ket{1,1} \rightarrow \ket{3}$. Then we can rewrite the set of states $\{\ket{a'_k},k=1,\cdots,12\}$ in $(\bbC^2)^{\otimes 4}$ as the set of bipartite product states $\{\ket{b_k}, k=1,\cdots,12\}$ in $\bbC^4 \otimes \bbC^4$ in Eq. \ref{eq:bk22}, that is,
		\begin{eqnarray}
			&&
			\ket{b_1}=\frac{1}{2\sqrt{2}}(\ket{1}+\ket{2})((1+i)\ket{1}+(1-i)\ket{2}), \notag \\
			&&
			\ket{b_2}=\frac{1}{2\sqrt{2}}((1+i)\ket{1}+(1-i)\ket{2})(\ket{0}-\ket{3}), \notag \\
			&&
			\ket{b_3}=\frac{1}{2\sqrt{2}}(\ket{0}-\ket{3})(\ket{0}-i\ket{2}+i\ket{1}-\ket{3}), \notag \\
			&&
			\ket{b_4}=\frac{1}{2\sqrt{2}}(\ket{0}+i\ket{1}-i\ket{2}-\ket{3})(\ket{1}+\ket{2}), \notag \\
			&&
			\ket{b_5}=\frac{1}{6}(\ket{0}+(1-i)\ket{1}+(1+i)\ket{2}-\ket{3})(\ket{0} \notag \\
			&&\qquad +(1-i)\ket{1}+(1+i)\ket{2}-\ket{3}), \notag \\
			&&
			\ket{b_6}=\frac{1}{2}(\ket{0}+\ket{3})(\ket{0}+\ket{3}), \notag \\
			&&
			\ket{b_7}=\frac{1}{2}(\ket{0}+\ket{3})(\ket{1}+\ket{2}), \notag \\ 
			&&
			\ket{b_8}=\frac{i}{2}(\ket{0}+\ket{3})(\ket{2}-\ket{1}), \notag \\
			&&
			\ket{b_9}=\frac{1}{2}(\ket{0}+\ket{3})(\ket{0}-\ket{3}), \notag \\
			&&
			\ket{b_{10}}=\frac{1}{2}(\ket{1}+\ket{2})(\ket{0}+\ket{3}), \notag \\
			&&
			\ket{b_{11}}=\frac{i}{2}(-\ket{1}+\ket{2})(\ket{0}+\ket{3}), \notag \\
			&&
			\ket{b_{12}}=\frac{1}{2}(\ket{0}-\ket{3})(\ket{0}+\ket{3}).
		\end{eqnarray} 
		
		The set of orthonormal states, $\{\frac{1}{\sqrt{2}}(\ket{0}-\ket{3}),\ket{1}, \ket{2} \}$, spans a three-dimensional Hilbert space $\cH_3$. Consider a bijection from $\cH_3$ to $\bbC^3$ as follows,
		\begin{eqnarray}
			\begin{aligned}
				\frac{1}{\sqrt{2}}(\ket{0}-\ket{3}) \in \cH_3 \qquad && \rightarrow \qquad \ket{0} \in \bbC^3,  \notag \\
				\ket{1} \in \cH_3 \qquad && \rightarrow \qquad \ket{1} \in \bbC^3,  \notag \\
				\ket{2}	\in \cH_3 \qquad && \rightarrow \qquad \ket{2} \in \bbC^3.
			\end{aligned}
		\end{eqnarray}
		Then $\ket{b_1}, \ket{b_2},\cdots,\ket{b_5} \in \cH_3 \otimes \cH_3$ can be transformed into $\ket{c_1}, \ket{c_2},\cdots,\ket{c_5} \in \bbC^3 \otimes \bbC^3$, where 
		\begin{eqnarray}
			\label{eq:upbiso}
			&&
			\ket{c_1}=\frac{1}{2\sqrt{2}}(\ket{1}+\ket{2})((1+i)\ket{1}+(1-i)\ket{2}), \notag \\
			&&
			\ket{c_2}=\frac{1}{2}((1+i)\ket{1}+(1-i)\ket{2})\ket{0}, \notag \\
			&&
			\ket{c_3}=\frac{1}{\sqrt{2}}\ket{0}(\ket{0}-i\ket{2}+i\ket{1}), \notag \\
			&&
			\ket{c_4}=\frac{1}{2}(\ket{0}+i\ket{1}-i\ket{2})(\ket{1}+\ket{2}), \notag \\
			&&
			\ket{c_5}=\frac{1}{3}(\ket{0}+(1-i)\ket{1}+(1+i)\ket{2})(\ket{0}+(1-i)\ket{1}+(1+i)\ket{2}). \notag \\
		\end{eqnarray} 
		Evidently, each partition of $\ket{c_1}, \ket{c_2},\cdots,\ket{c_5}$ are composed by the vectors $\ket{1}+\ket{2}, (1+i)\ket{1}+(1-i)\ket{2}, \ket{0}, \ket{0}+i\ket{1}-i\ket{2}$ and $\ket{0}+(1-i)\ket{1}+(1+i)\ket{2}$. One can show that any three of them are linearly independent. Thus using Lemma 1 in \cite{ISI:000184088700001}, we know that these five states $\ket{c_1}, \ket{c_2},\cdots,\ket{c_5}$ form a two-qutrit UPB. Thus the set $\{\ket{b_k}, k=1,\cdots,12\}$ in \eqref{eq:bk22} is locally indistinguishable. So the bipartite state $\ket{a'_k}$ in \eqref{eq:a_kspecial} are nonlocal in terms of the MES. Then from Definition \ref{def:nonlocaloperator}, one can obtain that the strongly UPUOB $\mathbb{U}_2$ is genuinely nonlocal in terms of the two level MES. This completes the proof.
	\end{proof}	
	
	Theorem \ref{thm:UPUOB22nonlocal} implies that one cannot distinguish the strongly UPUOB $\mathbb{U}_2$ in \eqref{eq:U1-5}-\eqref{eq:U6-12} in every bipartition using LOCC. Inspired by Theorem \ref{thm:UPUOB22nonlocal}, we ask whether a strongly UPUOB or UPUOB on higher dimensional Hilbert space is nonlocal or genuinely nonlocal. In the following theorem, we show that the strongly UPUOB $\mathbb{U}_2$ in \eqref{eq:U1-5}-\eqref{eq:U6-12} can be distinguished by the protocol which only consumes three ebits of entanglement resource.  
	
	\begin{theorem}
		\label{thm:ebitent}
		
		The strongly UPUOB $\mathbb{U}_2$ in \eqref{eq:U1-5}-\eqref{eq:U6-12} can be distinguished by three ebits of entanglement.
	\end{theorem}
	\begin{proof}
		From Theorem \ref{thm:UPUOB22nonlocal}, the strongly UPUOB $\mathbb{U}_2$ in \eqref{eq:U1-5}-\eqref{eq:U6-12} is genuinely nonlocal in terms of the two-level MES. To distinguish $\mathbb{U}_2$, it suffices to distinguish the states $\ket{a_k'}$'s \eqref{eq:ak22}. First, a 2-level MES is distributed between  $A_1$ and $B_1$, then $A_1$ teleports his subsystem to $B_1$ by using teleportation-based protocol \cite{bennett1993teleporting}. Next,  a 2-level MES is distributed between  $A_2$ and $B_2$, then $A_2$ also teleports his subsystem to $B_2$ by using teleportation-based protocol. It means that we obtain the set of bipartite product states $\{\ket{b_k},k=1,\cdots,12\}$ in $\cH_{A}\otimes \cH_{B}=\bbC^4 \otimes \bbC^4$, where
		\begin{eqnarray}
			\label{eq:bk22}
			&&
			\ket{b_1}=\frac{1}{2\sqrt{2}}(\ket{1}+\ket{2})((1+i)\ket{1}+(1-i)\ket{2}), \notag \\
			&&
			\ket{b_2}=\frac{1}{2\sqrt{2}}((1+i)\ket{1}+(1-i)\ket{2})(\ket{0}-\ket{3}), \notag \\
			&&
			\ket{b_3}=\frac{1}{2\sqrt{2}}(\ket{0}-\ket{3})(\ket{0}-i\ket{2}+i\ket{1}-\ket{3}), \notag \\
			&&
			\ket{b_4}=\frac{1}{2\sqrt{2}}(\ket{0}+i\ket{1}-i\ket{2}-\ket{3})(\ket{1}+\ket{2}), \notag \\
			&&
			\ket{b_5}=\frac{1}{6}(\ket{0}+(1-i)\ket{1}+(1+i)\ket{2}-\ket{3})(\ket{0} \notag \\
			&&\qquad +(1-i)\ket{1}+(1+i)\ket{2}-\ket{3}), \notag \\
			&&
			\ket{b_6}=\frac{1}{2}(\ket{0}+\ket{3})(\ket{0}+\ket{3}), \notag \\
			&&
			\ket{b_7}=\frac{1}{2}(\ket{0}+\ket{3})(\ket{1}+\ket{2}), \notag \\ 
			&&
			\ket{b_8}=\frac{i}{2}(\ket{0}+\ket{3})(\ket{2}-\ket{1}), \notag \\
			&&
			\ket{b_9}=\frac{1}{2}(\ket{0}+\ket{3})(\ket{0}-\ket{3}), \notag \\
			&&
			\ket{b_{10}}=\frac{1}{2}(\ket{1}+\ket{2})(\ket{0}+\ket{3}), \notag \\
			&&
			\ket{b_{11}}=\frac{i}{2}(-\ket{1}+\ket{2})(\ket{0}+\ket{3}), \notag \\
			&&
			\ket{b_{12}}=\frac{1}{2}(\ket{0}-\ket{3})(\ket{0}+\ket{3}).
		\end{eqnarray} 
		and $A=A_1B_1$, $B=A_2B_2$. Here it consumes two ebits of entanglement resource. Now the discrimination protocol proceeds as follows.
		
		\emph{Step 1}. Alice performs the measurement $\{M_1:=\frac{1}{2}(\ket{0}+\ket{3})(\bra{0}+\bra{3})_A, \overline{M}_1=I-M_1\}$. If $M_1$ clicks, Eq. \eqref{eq:bk22} remains $\{ \ket{b_k}\}_{k=6}^9$. These four states can be easily distinguished by Bob \cite{PhysRevLett.85.4972}. If $\overline{M}_1$ clicks, Eq. \eqref{eq:bk22} remains $\{\ket{b_k}\}_{k=1}^5 \cup \{\ket{b_k}\}_{k=10}^{12}$.
		
		\emph{Step 2}. Bob performs the measurement $\{M_2:=\frac{1}{2}(\ket{0}+\ket{3})(\bra{0}+\bra{3})_B, \overline{M}_2=I-M_2\}$. If $M_2$ clicks, $\{\ket{b_k}\}_{k=1}^5 \cup \{\ket{b_k}\}_{k=10}^{12}$ remains $\ket{b_{10}}, \ket{b_{11}}, \ket{b_{12}}$. These three states can be easily distinguished by Alice. If $\overline{M}_2$ clicks, $\{\ket{b_k}\}_{k=1}^5 \cup \{\ket{b_k}\}_{k=10}^{12}$ remains $\{\ket{b_k}\}_{k=1}^5$.
		
		\emph{Step 3}. In the proof of Theorem \ref{thm:UPUOB22nonlocal}, we have shown that $\ket{b_1},\cdots,\ket{b_5}$ is isomorphic to a two-qutrit UPB $\ket{c_1},\ket{c_2},\cdots,\ket{c_5}$ in Eq. \eqref{eq:upbiso}. It has been proved that one ebit of entanglement is sufficient to distinguish any UPB on $3 \otimes 3$ only by LOCC \cite{PhysRevA.77.012304}. 	That is, a 2-level MES is distributed between Alice and Bob. The initial state is 
		\begin{eqnarray}
			\label{eq:ebitbk22}
			\ket{\phi_k}=\ket{c_k}_{AB}\otimes(\ket{00}_{ab}+\ket{11}_{ab}),
		\end{eqnarray}
		where $a,b$ are the ancillary systems of Alice and Bob, respectively, $k=1,2,\cdots,5$. Thus these five states $\{\ket{b_k}\}_{k=1}^5$ can be locally distinguished. 
		
		So it requires three ebits of entanglement resource to distinguish the strongly UPUOB $\mathbb{U}_2$ in \eqref{eq:U1-5}-\eqref{eq:U6-12}.
	\end{proof}
	
	Since a strongly UPUOB cannot be locally distinguished in every bipartition, a perfect entanglement-assisted discrimination with less entanglement resource is desirable. As an application of Theorem \ref{thm:ebitent}, FIG. \ref{fig:discriminationU2} shows that it requires three ebits of entanglement resource to distinguish the strongly UPUOB $\mathbb{U}_2$ in \eqref{eq:U1-5}-\eqref{eq:U6-12} which was applied to $A_1$ and $A_2$. Specifically, one two-level MES is distributed between $A_1$ and $B_1$, and one two-level MES is distributed between $A_2$ and $B_2$. It consumes a two-level MES to distinguish the strongly UPUOB $\mathbb{U}_2$ in \eqref{eq:U1-5}-\eqref{eq:U6-12} between systems $a$ and $b$, where $a,b$ are the ancillary systems of Alice and Bob, respectively. Since the simplest type of product unitary gates are of the general form $U_i=P_i \otimes V_i$ acting on a bipartite Hilbert space $\mathcal{H}_A \otimes \mathcal{H}_B$, where $P_i$'s are orthogonal unitary operators on $\mathcal{H}_A$ and $V_i$'s are orthogonal unitary operators on $\mathcal{H}_B$. One can find that the strongly UPUOB $\mathbb{U}_2$ in \eqref{eq:U1-5}-\eqref{eq:U6-12} is one of the simplest product unitary operators which are operational in experiments.  Thus the two-dimensional strongly UPUOB can be used in multipartite secret sharing \cite{PhysRevLett.126.210505, shamir1979share}. For more complicated product unitary gates discrimination and secret sharing scheme, we need to find multipartite strongly UPUOB in higher dimension.     
	
	\section{Conclusions}
\label{sec:con}
Inspired by the notion of UPB, we have proposed the definitions of UPOB, UPUOB, strongly UPUOB as well as the nonlocality and genuinely nonlocality of unitary operators. We have presented the properties and examples of them. Further, we have investigated the nonlocality of the two-qubit strongly UPUOB in terms of MES. There are many interesting problems left. We do not know any construction of strongly UPUOB in more general cases. Are UPUOBs and strongly UPUOBs nonlocal or genuinely nonlocal in terms of MES? Can we find other bipartite state $\ket{\psi}$ rather than MES such that the states $\ket{a_k}$ in Eq. \eqref{eq:a_k} are locally indistinguishable across any bipartition? 

\section*{Acknowledgments}
M.H. and L.C. were supported by the NNSF of China (Grant No. 11871089), and the Fundamental Research Funds for the Central Universities (Grant No. ZG216S2110).  F.S. and X.Z. were supported by the NSFC under Grants No. 12171452 and No. 11771419, the Anhui Initiative in Quantum Information Technologies under Grant No. AHY150200, the Innovation Program for Quantum Science and Technology (2021ZD0302904), and the National Key Research and Development Program of China (2020YFA0713100). This work has received support from the European Union's Horizon Europe program through the ERC StG FINE-TEA-SQUAD (Grant No. 101040729).  This publication is part of the `Quantum Inspire – the Dutch Quantum Computer in the Cloud' project (with project number [NWA.1292.19.194]) of the NWA research program `Research on Routes by Consortia (ORC)', which is funded by the Netherlands Organization for Scientific Research (NWO).

\nocite{*}
\bibliography{mengyao=dis}

\begin{thebibliography}{39}%
\makeatletter
\providecommand \@ifxundefined [1]{%
 \@ifx{#1\undefined}
}%
\providecommand \@ifnum [1]{%
 \ifnum #1\expandafter \@firstoftwo
 \else \expandafter \@secondoftwo
 \fi
}%
\providecommand \@ifx [1]{%
 \ifx #1\expandafter \@firstoftwo
 \else \expandafter \@secondoftwo
 \fi
}%
\providecommand \natexlab [1]{#1}%
\providecommand \enquote  [1]{``#1''}%
\providecommand \bibnamefont  [1]{#1}%
\providecommand \bibfnamefont [1]{#1}%
\providecommand \citenamefont [1]{#1}%
\providecommand \href@noop [0]{\@secondoftwo}%
\providecommand \href [0]{\begingroup \@sanitize@url \@href}%
\providecommand \@href[1]{\@@startlink{#1}\@@href}%
\providecommand \@@href[1]{\endgroup#1\@@endlink}%
\providecommand \@sanitize@url [0]{\catcode `\\12\catcode `\$12\catcode
  `\&12\catcode `\#12\catcode `\^12\catcode `\_12\catcode `\%12\relax}%
\providecommand \@@startlink[1]{}%
\providecommand \@@endlink[0]{}%
\providecommand \url  [0]{\begingroup\@sanitize@url \@url }%
\providecommand \@url [1]{\endgroup\@href {#1}{\urlprefix }}%
\providecommand \urlprefix  [0]{URL }%
\providecommand \Eprint [0]{\href }%
\providecommand \doibase [0]{http://dx.doi.org/}%
\providecommand \selectlanguage [0]{\@gobble}%
\providecommand \bibinfo  [0]{\@secondoftwo}%
\providecommand \bibfield  [0]{\@secondoftwo}%
\providecommand \translation [1]{[#1]}%
\providecommand \BibitemOpen [0]{}%
\providecommand \bibitemStop [0]{}%
\providecommand \bibitemNoStop [0]{.\EOS\space}%
\providecommand \EOS [0]{\spacefactor3000\relax}%
\providecommand \BibitemShut  [1]{\csname bibitem#1\endcsname}%
\let\auto@bib@innerbib\@empty
\bibitem [{\citenamefont {Shi}\ \emph {et~al.}(2020{\natexlab{a}})\citenamefont
  {Shi}, \citenamefont {Hu}, \citenamefont {Chen},\ and\ \citenamefont
  {Zhang}}]{PhysRevA.102.042202}%
  \BibitemOpen
  \bibfield  {author} {\bibinfo {author} {\bibfnamefont {F.}~\bibnamefont
  {Shi}}, \bibinfo {author} {\bibfnamefont {M.}~\bibnamefont {Hu}}, \bibinfo
  {author} {\bibfnamefont {L.}~\bibnamefont {Chen}}, \ and\ \bibinfo {author}
  {\bibfnamefont {X.}~\bibnamefont {Zhang}},\ }\href {\doibase
  10.1103/PhysRevA.102.042202} {\bibfield  {journal} {\bibinfo  {journal}
  {Phys. Rev. A}\ }\textbf {\bibinfo {volume} {102}},\ \bibinfo {pages}
  {042202} (\bibinfo {year} {2020}{\natexlab{a}})}\BibitemShut {NoStop}%
\bibitem [{\citenamefont {Halder}\ \emph {et~al.}(2019)\citenamefont {Halder},
  \citenamefont {Banik}, \citenamefont {Agrawal},\ and\ \citenamefont
  {Bandyopadhyay}}]{PhysRevLett.122.040403}%
  \BibitemOpen
  \bibfield  {author} {\bibinfo {author} {\bibfnamefont {S.}~\bibnamefont
  {Halder}}, \bibinfo {author} {\bibfnamefont {M.}~\bibnamefont {Banik}},
  \bibinfo {author} {\bibfnamefont {S.}~\bibnamefont {Agrawal}}, \ and\
  \bibinfo {author} {\bibfnamefont {S.}~\bibnamefont {Bandyopadhyay}},\ }\href
  {\doibase 10.1103/PhysRevLett.122.040403} {\bibfield  {journal} {\bibinfo
  {journal} {Phys. Rev. Lett.}\ }\textbf {\bibinfo {volume} {122}},\ \bibinfo
  {pages} {040403} (\bibinfo {year} {2019})}\BibitemShut {NoStop}%
\bibitem [{\citenamefont {Bennett}\ \emph
  {et~al.}(1999{\natexlab{a}})\citenamefont {Bennett}, \citenamefont
  {DiVincenzo}, \citenamefont {Fuchs}, \citenamefont {Mor}, \citenamefont
  {Rains}, \citenamefont {Shor}, \citenamefont {Smolin},\ and\ \citenamefont
  {Wootters}}]{PhysRevA.59.1070}%
  \BibitemOpen
  \bibfield  {author} {\bibinfo {author} {\bibfnamefont {C.~H.}\ \bibnamefont
  {Bennett}}, \bibinfo {author} {\bibfnamefont {D.~P.}\ \bibnamefont
  {DiVincenzo}}, \bibinfo {author} {\bibfnamefont {C.~A.}\ \bibnamefont
  {Fuchs}}, \bibinfo {author} {\bibfnamefont {T.}~\bibnamefont {Mor}}, \bibinfo
  {author} {\bibfnamefont {E.}~\bibnamefont {Rains}}, \bibinfo {author}
  {\bibfnamefont {P.~W.}\ \bibnamefont {Shor}}, \bibinfo {author}
  {\bibfnamefont {J.~A.}\ \bibnamefont {Smolin}}, \ and\ \bibinfo {author}
  {\bibfnamefont {W.~K.}\ \bibnamefont {Wootters}},\ }\href {\doibase
  10.1103/PhysRevA.59.1070} {\bibfield  {journal} {\bibinfo  {journal} {Phys.
  Rev. A}\ }\textbf {\bibinfo {volume} {59}},\ \bibinfo {pages} {1070}
  (\bibinfo {year} {1999}{\natexlab{a}})}\BibitemShut {NoStop}%
\bibitem [{\citenamefont {Deng}(2018)}]{PhysRevLett.120.240402}%
  \BibitemOpen
  \bibfield  {author} {\bibinfo {author} {\bibfnamefont {D.-L.}\ \bibnamefont
  {Deng}},\ }\href {\doibase 10.1103/PhysRevLett.120.240402} {\bibfield
  {journal} {\bibinfo  {journal} {Phys. Rev. Lett.}\ }\textbf {\bibinfo
  {volume} {120}},\ \bibinfo {pages} {240402} (\bibinfo {year}
  {2018})}\BibitemShut {NoStop}%
\bibitem [{\citenamefont {Liang}\ \emph {et~al.}(2014)\citenamefont {Liang},
  \citenamefont {Curchod}, \citenamefont {Bowles},\ and\ \citenamefont
  {Gisin}}]{PhysRevLett.113.130401}%
  \BibitemOpen
  \bibfield  {author} {\bibinfo {author} {\bibfnamefont {Y.-C.}\ \bibnamefont
  {Liang}}, \bibinfo {author} {\bibfnamefont {F.~J.}\ \bibnamefont {Curchod}},
  \bibinfo {author} {\bibfnamefont {J.}~\bibnamefont {Bowles}}, \ and\ \bibinfo
  {author} {\bibfnamefont {N.}~\bibnamefont {Gisin}},\ }\href {\doibase
  10.1103/PhysRevLett.113.130401} {\bibfield  {journal} {\bibinfo  {journal}
  {Phys. Rev. Lett.}\ }\textbf {\bibinfo {volume} {113}},\ \bibinfo {pages}
  {130401} (\bibinfo {year} {2014})}\BibitemShut {NoStop}%
\bibitem [{\citenamefont {Cohen}(2008)}]{PhysRevA.77.012304}%
  \BibitemOpen
  \bibfield  {author} {\bibinfo {author} {\bibfnamefont {S.~M.}\ \bibnamefont
  {Cohen}},\ }\href {\doibase 10.1103/PhysRevA.77.012304} {\bibfield  {journal}
  {\bibinfo  {journal} {Phys. Rev. A}\ }\textbf {\bibinfo {volume} {77}},\
  \bibinfo {pages} {012304} (\bibinfo {year} {2008})}\BibitemShut {NoStop}%
\bibitem [{\citenamefont {DiVincenzo}\ \emph {et~al.}(2003)\citenamefont
  {DiVincenzo}, \citenamefont {Mor}, \citenamefont {Shor}, \citenamefont
  {Smolin},\ and\ \citenamefont {Terhal}}]{ISI:000184088700001}%
  \BibitemOpen
  \bibfield  {author} {\bibinfo {author} {\bibfnamefont {D.}~\bibnamefont
  {DiVincenzo}}, \bibinfo {author} {\bibfnamefont {T.}~\bibnamefont {Mor}},
  \bibinfo {author} {\bibfnamefont {P.}~\bibnamefont {Shor}}, \bibinfo {author}
  {\bibfnamefont {J.}~\bibnamefont {Smolin}}, \ and\ \bibinfo {author}
  {\bibfnamefont {B.}~\bibnamefont {Terhal}},\ }\href {\doibase
  10.1007/s00220-003-0877-6} {\bibfield  {journal} {\bibinfo  {journal}
  {Communications in Mathematical Physics}\ }\textbf {\bibinfo {volume}
  {238}},\ \bibinfo {pages} {379} (\bibinfo {year} {2003})}\BibitemShut
  {NoStop}%
\bibitem [{\citenamefont {De~Rinaldis}(2004)}]{PhysRevA.70.022309}%
  \BibitemOpen
  \bibfield  {author} {\bibinfo {author} {\bibfnamefont {S.}~\bibnamefont
  {De~Rinaldis}},\ }\href {\doibase 10.1103/PhysRevA.70.022309} {\bibfield
  {journal} {\bibinfo  {journal} {Phys. Rev. A}\ }\textbf {\bibinfo {volume}
  {70}},\ \bibinfo {pages} {022309} (\bibinfo {year} {2004})}\BibitemShut
  {NoStop}%
\bibitem [{\citenamefont {Lanyon}\ \emph {et~al.}(2009)\citenamefont {Lanyon},
  \citenamefont {Barbieri}, \citenamefont {Almeida}, \citenamefont {Jennewein},
  \citenamefont {Ralph}, \citenamefont {Resch}, \citenamefont {Pryde},
  \citenamefont {O’brien}, \citenamefont {Gilchrist},\ and\ \citenamefont
  {White}}]{lanyon2009simplifying}%
  \BibitemOpen
  \bibfield  {author} {\bibinfo {author} {\bibfnamefont {B.~P.}\ \bibnamefont
  {Lanyon}}, \bibinfo {author} {\bibfnamefont {M.}~\bibnamefont {Barbieri}},
  \bibinfo {author} {\bibfnamefont {M.~P.}\ \bibnamefont {Almeida}}, \bibinfo
  {author} {\bibfnamefont {T.}~\bibnamefont {Jennewein}}, \bibinfo {author}
  {\bibfnamefont {T.~C.}\ \bibnamefont {Ralph}}, \bibinfo {author}
  {\bibfnamefont {K.~J.}\ \bibnamefont {Resch}}, \bibinfo {author}
  {\bibfnamefont {G.~J.}\ \bibnamefont {Pryde}}, \bibinfo {author}
  {\bibfnamefont {J.~L.}\ \bibnamefont {O’brien}}, \bibinfo {author}
  {\bibfnamefont {A.}~\bibnamefont {Gilchrist}}, \ and\ \bibinfo {author}
  {\bibfnamefont {A.~G.}\ \bibnamefont {White}},\ }\href
  {https://www.nature.com/articles/nphys1150} {\bibfield  {journal} {\bibinfo
  {journal} {Nature Physics}\ }\textbf {\bibinfo {volume} {5}},\ \bibinfo
  {pages} {134} (\bibinfo {year} {2009})}\BibitemShut {NoStop}%
\bibitem [{\citenamefont {Procopio}\ \emph {et~al.}(2015)\citenamefont
  {Procopio}, \citenamefont {Moqanaki}, \citenamefont {Ara{\'u}jo},
  \citenamefont {Costa}, \citenamefont {Calafell}, \citenamefont {Dowd},
  \citenamefont {Hamel}, \citenamefont {Rozema}, \citenamefont {Brukner},\ and\
  \citenamefont {Walther}}]{procopio2015experimental}%
  \BibitemOpen
  \bibfield  {author} {\bibinfo {author} {\bibfnamefont {L.~M.}\ \bibnamefont
  {Procopio}}, \bibinfo {author} {\bibfnamefont {A.}~\bibnamefont {Moqanaki}},
  \bibinfo {author} {\bibfnamefont {M.}~\bibnamefont {Ara{\'u}jo}}, \bibinfo
  {author} {\bibfnamefont {F.}~\bibnamefont {Costa}}, \bibinfo {author}
  {\bibfnamefont {I.~A.}\ \bibnamefont {Calafell}}, \bibinfo {author}
  {\bibfnamefont {E.~G.}\ \bibnamefont {Dowd}}, \bibinfo {author}
  {\bibfnamefont {D.~R.}\ \bibnamefont {Hamel}}, \bibinfo {author}
  {\bibfnamefont {L.~A.}\ \bibnamefont {Rozema}}, \bibinfo {author}
  {\bibfnamefont {{\v{C}}.}~\bibnamefont {Brukner}}, \ and\ \bibinfo {author}
  {\bibfnamefont {P.}~\bibnamefont {Walther}},\ }\href
  {https://www.nature.com/articles/ncomms8913#citeas} {\bibfield  {journal}
  {\bibinfo  {journal} {Nature communications}\ }\textbf {\bibinfo {volume}
  {6}},\ \bibinfo {pages} {1} (\bibinfo {year} {2015})}\BibitemShut {NoStop}%
\bibitem [{\citenamefont {Bagchi}\ and\ \citenamefont
  {Pati}(2016)}]{bagchi2016uncertainty}%
  \BibitemOpen
  \bibfield  {author} {\bibinfo {author} {\bibfnamefont {S.}~\bibnamefont
  {Bagchi}}\ and\ \bibinfo {author} {\bibfnamefont {A.~K.}\ \bibnamefont
  {Pati}},\ }\href {\doibase 10.1103/PhysRevA.94.042104} {\bibfield  {journal}
  {\bibinfo  {journal} {Phys. Rev. A}\ }\textbf {\bibinfo {volume} {94}},\
  \bibinfo {pages} {042104} (\bibinfo {year} {2016})}\BibitemShut {NoStop}%
\bibitem [{\citenamefont {Chen}\ and\ \citenamefont
  {Yu}(2014)}]{PhysRevA.89.062326}%
  \BibitemOpen
  \bibfield  {author} {\bibinfo {author} {\bibfnamefont {L.}~\bibnamefont
  {Chen}}\ and\ \bibinfo {author} {\bibfnamefont {L.}~\bibnamefont {Yu}},\
  }\href {\doibase 10.1103/PhysRevA.89.062326} {\bibfield  {journal} {\bibinfo
  {journal} {Phys. Rev. A}\ }\textbf {\bibinfo {volume} {89}},\ \bibinfo
  {pages} {062326} (\bibinfo {year} {2014})}\BibitemShut {NoStop}%
\bibitem [{\citenamefont {Eisert}\ \emph {et~al.}(2000)\citenamefont {Eisert},
  \citenamefont {Jacobs}, \citenamefont {Papadopoulos},\ and\ \citenamefont
  {Plenio}}]{PhysRevA.62.052317}%
  \BibitemOpen
  \bibfield  {author} {\bibinfo {author} {\bibfnamefont {J.}~\bibnamefont
  {Eisert}}, \bibinfo {author} {\bibfnamefont {K.}~\bibnamefont {Jacobs}},
  \bibinfo {author} {\bibfnamefont {P.}~\bibnamefont {Papadopoulos}}, \ and\
  \bibinfo {author} {\bibfnamefont {M.~B.}\ \bibnamefont {Plenio}},\ }\href
  {\doibase 10.1103/PhysRevA.62.052317} {\bibfield  {journal} {\bibinfo
  {journal} {Phys. Rev. A}\ }\textbf {\bibinfo {volume} {62}},\ \bibinfo
  {pages} {052317} (\bibinfo {year} {2000})}\BibitemShut {NoStop}%
\bibitem [{\citenamefont {Cavalcanti}\ \emph {et~al.}(2011)\citenamefont
  {Cavalcanti}, \citenamefont {Almeida}, \citenamefont {Scarani},\ and\
  \citenamefont {Acin}}]{cite-key}%
  \BibitemOpen
  \bibfield  {author} {\bibinfo {author} {\bibfnamefont {D.}~\bibnamefont
  {Cavalcanti}}, \bibinfo {author} {\bibfnamefont {M.~L.}\ \bibnamefont
  {Almeida}}, \bibinfo {author} {\bibfnamefont {V.}~\bibnamefont {Scarani}}, \
  and\ \bibinfo {author} {\bibfnamefont {A.}~\bibnamefont {Acin}},\ }\href
  {\doibase 10.1038/ncomms1193} {\bibfield  {journal} {\bibinfo  {journal}
  {Nature Communications}\ }\textbf {\bibinfo {volume} {2}},\ \bibinfo {pages}
  {184} (\bibinfo {year} {2011})}\BibitemShut {NoStop}%
\bibitem [{\citenamefont {Brunner}\ \emph {et~al.}(2014)\citenamefont
  {Brunner}, \citenamefont {Cavalcanti}, \citenamefont {Pironio}, \citenamefont
  {Scarani},\ and\ \citenamefont {Wehner}}]{RevModPhys.86.419}%
  \BibitemOpen
  \bibfield  {author} {\bibinfo {author} {\bibfnamefont {N.}~\bibnamefont
  {Brunner}}, \bibinfo {author} {\bibfnamefont {D.}~\bibnamefont {Cavalcanti}},
  \bibinfo {author} {\bibfnamefont {S.}~\bibnamefont {Pironio}}, \bibinfo
  {author} {\bibfnamefont {V.}~\bibnamefont {Scarani}}, \ and\ \bibinfo
  {author} {\bibfnamefont {S.}~\bibnamefont {Wehner}},\ }\href {\doibase
  10.1103/RevModPhys.86.419} {\bibfield  {journal} {\bibinfo  {journal} {Rev.
  Mod. Phys.}\ }\textbf {\bibinfo {volume} {86}},\ \bibinfo {pages} {419}
  (\bibinfo {year} {2014})}\BibitemShut {NoStop}%
\bibitem [{\citenamefont {Renou}\ \emph {et~al.}(2019)\citenamefont {Renou},
  \citenamefont {B\"aumer}, \citenamefont {Boreiri}, \citenamefont {Brunner},
  \citenamefont {Gisin},\ and\ \citenamefont {Beigi}}]{PhysRevLett.123.140401}%
  \BibitemOpen
  \bibfield  {author} {\bibinfo {author} {\bibfnamefont {M.-O.}\ \bibnamefont
  {Renou}}, \bibinfo {author} {\bibfnamefont {E.}~\bibnamefont {B\"aumer}},
  \bibinfo {author} {\bibfnamefont {S.}~\bibnamefont {Boreiri}}, \bibinfo
  {author} {\bibfnamefont {N.}~\bibnamefont {Brunner}}, \bibinfo {author}
  {\bibfnamefont {N.}~\bibnamefont {Gisin}}, \ and\ \bibinfo {author}
  {\bibfnamefont {S.}~\bibnamefont {Beigi}},\ }\href {\doibase
  10.1103/PhysRevLett.123.140401} {\bibfield  {journal} {\bibinfo  {journal}
  {Phys. Rev. Lett.}\ }\textbf {\bibinfo {volume} {123}},\ \bibinfo {pages}
  {140401} (\bibinfo {year} {2019})}\BibitemShut {NoStop}%
\bibitem [{\citenamefont {Banik}\ \emph {et~al.}(2021)\citenamefont {Banik},
  \citenamefont {Guha}, \citenamefont {Alimuddin}, \citenamefont {Kar},
  \citenamefont {Halder},\ and\ \citenamefont
  {Bhattacharya}}]{PhysRevLett.126.210505}%
  \BibitemOpen
  \bibfield  {author} {\bibinfo {author} {\bibfnamefont {M.}~\bibnamefont
  {Banik}}, \bibinfo {author} {\bibfnamefont {T.}~\bibnamefont {Guha}},
  \bibinfo {author} {\bibfnamefont {M.}~\bibnamefont {Alimuddin}}, \bibinfo
  {author} {\bibfnamefont {G.}~\bibnamefont {Kar}}, \bibinfo {author}
  {\bibfnamefont {S.}~\bibnamefont {Halder}}, \ and\ \bibinfo {author}
  {\bibfnamefont {S.~S.}\ \bibnamefont {Bhattacharya}},\ }\href {\doibase
  10.1103/PhysRevLett.126.210505} {\bibfield  {journal} {\bibinfo  {journal}
  {Phys. Rev. Lett.}\ }\textbf {\bibinfo {volume} {126}},\ \bibinfo {pages}
  {210505} (\bibinfo {year} {2021})}\BibitemShut {NoStop}%
\bibitem [{\citenamefont {Reck}\ \emph {et~al.}(1994)\citenamefont {Reck},
  \citenamefont {Zeilinger}, \citenamefont {Bernstein},\ and\ \citenamefont
  {Bertani}}]{reck1994experimental}%
  \BibitemOpen
  \bibfield  {author} {\bibinfo {author} {\bibfnamefont {M.}~\bibnamefont
  {Reck}}, \bibinfo {author} {\bibfnamefont {A.}~\bibnamefont {Zeilinger}},
  \bibinfo {author} {\bibfnamefont {H.~J.}\ \bibnamefont {Bernstein}}, \ and\
  \bibinfo {author} {\bibfnamefont {P.}~\bibnamefont {Bertani}},\ }\href@noop
  {} {\bibfield  {journal} {\bibinfo  {journal} {Physical review letters}\
  }\textbf {\bibinfo {volume} {73}},\ \bibinfo {pages} {58} (\bibinfo {year}
  {1994})}\BibitemShut {NoStop}%
\bibitem [{\citenamefont {Usher}\ \emph {et~al.}(2017)\citenamefont {Usher},
  \citenamefont {Hoban},\ and\ \citenamefont {Browne}}]{PhysRevA.96.032321}%
  \BibitemOpen
  \bibfield  {author} {\bibinfo {author} {\bibfnamefont {N.}~\bibnamefont
  {Usher}}, \bibinfo {author} {\bibfnamefont {M.~J.}\ \bibnamefont {Hoban}}, \
  and\ \bibinfo {author} {\bibfnamefont {D.~E.}\ \bibnamefont {Browne}},\
  }\href {\doibase 10.1103/PhysRevA.96.032321} {\bibfield  {journal} {\bibinfo
  {journal} {Phys. Rev. A}\ }\textbf {\bibinfo {volume} {96}},\ \bibinfo
  {pages} {032321} (\bibinfo {year} {2017})}\BibitemShut {NoStop}%
\bibitem [{\citenamefont {Zhou}\ and\ \citenamefont {Ying}(2017)}]{8049724}%
  \BibitemOpen
  \bibfield  {author} {\bibinfo {author} {\bibfnamefont {L.}~\bibnamefont
  {Zhou}}\ and\ \bibinfo {author} {\bibfnamefont {M.}~\bibnamefont {Ying}},\
  }in\ \href {\doibase 10.1109/CSF.2017.23} {\emph {\bibinfo {booktitle} {2017
  IEEE 30th Computer Security Foundations Symposium (CSF)}}}\ (\bibinfo {year}
  {2017})\ pp.\ \bibinfo {pages} {249--262}\BibitemShut {NoStop}%
\bibitem [{\citenamefont {Song}\ \emph {et~al.}(2019)\citenamefont {Song},
  \citenamefont {Cui}, \citenamefont {Wang}, \citenamefont {Hao}, \citenamefont
  {Feng},\ and\ \citenamefont {Li}}]{song2019quantum}%
  \BibitemOpen
  \bibfield  {author} {\bibinfo {author} {\bibfnamefont {C.}~\bibnamefont
  {Song}}, \bibinfo {author} {\bibfnamefont {J.}~\bibnamefont {Cui}}, \bibinfo
  {author} {\bibfnamefont {H.}~\bibnamefont {Wang}}, \bibinfo {author}
  {\bibfnamefont {J.}~\bibnamefont {Hao}}, \bibinfo {author} {\bibfnamefont
  {H.}~\bibnamefont {Feng}}, \ and\ \bibinfo {author} {\bibfnamefont
  {Y.}~\bibnamefont {Li}},\ }\href {\doibase 10.1126/sciadv.aaw5686} {\bibfield
   {journal} {\bibinfo  {journal} {Science Advances}\ }\textbf {\bibinfo
  {volume} {5}},\ \bibinfo {pages} {eaaw5686} (\bibinfo {year} {2019})},\
  \Eprint
  {http://arxiv.org/abs/https://www.science.org/doi/pdf/10.1126/sciadv.aaw5686}
  {https://www.science.org/doi/pdf/10.1126/sciadv.aaw5686} \BibitemShut
  {NoStop}%
\bibitem [{\citenamefont {Hu}\ and\ \citenamefont
  {Chen}(2022)}]{hu2021genuine}%
  \BibitemOpen
  \bibfield  {author} {\bibinfo {author} {\bibfnamefont {M.}~\bibnamefont
  {Hu}}\ and\ \bibinfo {author} {\bibfnamefont {L.}~\bibnamefont {Chen}},\
  }\href {\doibase 10.1007/s11128-022-03497-7} {\bibfield  {journal} {\bibinfo
  {journal} {Quantum Information Processing}\ }\textbf {\bibinfo {volume}
  {21}},\ \bibinfo {pages} {162} (\bibinfo {year} {2022})}\BibitemShut
  {NoStop}%
\bibitem [{\citenamefont {Tian}\ \emph {et~al.}(2015)\citenamefont {Tian},
  \citenamefont {Yu}, \citenamefont {Gao}, \citenamefont {Wen},\ and\
  \citenamefont {Oh}}]{PhysRevA.91.052314}%
  \BibitemOpen
  \bibfield  {author} {\bibinfo {author} {\bibfnamefont {G.}~\bibnamefont
  {Tian}}, \bibinfo {author} {\bibfnamefont {S.}~\bibnamefont {Yu}}, \bibinfo
  {author} {\bibfnamefont {F.}~\bibnamefont {Gao}}, \bibinfo {author}
  {\bibfnamefont {Q.}~\bibnamefont {Wen}}, \ and\ \bibinfo {author}
  {\bibfnamefont {C.~H.}\ \bibnamefont {Oh}},\ }\href {\doibase
  10.1103/PhysRevA.91.052314} {\bibfield  {journal} {\bibinfo  {journal} {Phys.
  Rev. A}\ }\textbf {\bibinfo {volume} {91}},\ \bibinfo {pages} {052314}
  (\bibinfo {year} {2015})}\BibitemShut {NoStop}%
\bibitem [{\citenamefont {Yu}\ \emph {et~al.}(2012)\citenamefont {Yu},
  \citenamefont {Duan},\ and\ \citenamefont {Ying}}]{PhysRevLett.109.020506}%
  \BibitemOpen
  \bibfield  {author} {\bibinfo {author} {\bibfnamefont {N.}~\bibnamefont
  {Yu}}, \bibinfo {author} {\bibfnamefont {R.}~\bibnamefont {Duan}}, \ and\
  \bibinfo {author} {\bibfnamefont {M.}~\bibnamefont {Ying}},\ }\href {\doibase
  10.1103/PhysRevLett.109.020506} {\bibfield  {journal} {\bibinfo  {journal}
  {Phys. Rev. Lett.}\ }\textbf {\bibinfo {volume} {109}},\ \bibinfo {pages}
  {020506} (\bibinfo {year} {2012})}\BibitemShut {NoStop}%
\bibitem [{\citenamefont {Fan}(2004)}]{PhysRevLett.92.177905}%
  \BibitemOpen
  \bibfield  {author} {\bibinfo {author} {\bibfnamefont {H.}~\bibnamefont
  {Fan}},\ }\href {\doibase 10.1103/PhysRevLett.92.177905} {\bibfield
  {journal} {\bibinfo  {journal} {Phys. Rev. Lett.}\ }\textbf {\bibinfo
  {volume} {92}},\ \bibinfo {pages} {177905} (\bibinfo {year}
  {2004})}\BibitemShut {NoStop}%
\bibitem [{\citenamefont {Matthews}\ \emph {et~al.}(2009)\citenamefont
  {Matthews}, \citenamefont {Wehner},\ and\ \citenamefont
  {Winter}}]{matthews2009distinguishability}%
  \BibitemOpen
  \bibfield  {author} {\bibinfo {author} {\bibfnamefont {W.}~\bibnamefont
  {Matthews}}, \bibinfo {author} {\bibfnamefont {S.}~\bibnamefont {Wehner}}, \
  and\ \bibinfo {author} {\bibfnamefont {A.}~\bibnamefont {Winter}},\ }\href
  {https://link.springer.com/article/10.1007/s00220-009-0890-5} {\bibfield
  {journal} {\bibinfo  {journal} {Communications in Mathematical Physics}\
  }\textbf {\bibinfo {volume} {291}},\ \bibinfo {pages} {813} (\bibinfo {year}
  {2009})}\BibitemShut {NoStop}%
\bibitem [{\citenamefont {DiVincenzo}\ \emph {et~al.}(2002)\citenamefont
  {DiVincenzo}, \citenamefont {Leung},\ and\ \citenamefont
  {Terhal}}]{divincenzo2001quantum}%
  \BibitemOpen
  \bibfield  {author} {\bibinfo {author} {\bibfnamefont {D.}~\bibnamefont
  {DiVincenzo}}, \bibinfo {author} {\bibfnamefont {D.}~\bibnamefont {Leung}}, \
  and\ \bibinfo {author} {\bibfnamefont {B.}~\bibnamefont {Terhal}},\ }\href
  {\doibase 10.1109/18.985948} {\bibfield  {journal} {\bibinfo  {journal} {IEEE
  Transactions on Information Theory}\ }\textbf {\bibinfo {volume} {48}},\
  \bibinfo {pages} {580} (\bibinfo {year} {2002})}\BibitemShut {NoStop}%
\bibitem [{\citenamefont {Markham}\ and\ \citenamefont
  {Sanders}(2008)}]{PhysRevA.78.042309}%
  \BibitemOpen
  \bibfield  {author} {\bibinfo {author} {\bibfnamefont {D.}~\bibnamefont
  {Markham}}\ and\ \bibinfo {author} {\bibfnamefont {B.~C.}\ \bibnamefont
  {Sanders}},\ }\href {\doibase 10.1103/PhysRevA.78.042309} {\bibfield
  {journal} {\bibinfo  {journal} {Phys. Rev. A}\ }\textbf {\bibinfo {volume}
  {78}},\ \bibinfo {pages} {042309} (\bibinfo {year} {2008})}\BibitemShut
  {NoStop}%
\bibitem [{\citenamefont {Bennett}\ \emph
  {et~al.}(1999{\natexlab{b}})\citenamefont {Bennett}, \citenamefont
  {DiVincenzo}, \citenamefont {Mor}, \citenamefont {Shor}, \citenamefont
  {Smolin},\ and\ \citenamefont {Terhal}}]{PhysRevLett.82.5385}%
  \BibitemOpen
  \bibfield  {author} {\bibinfo {author} {\bibfnamefont {C.~H.}\ \bibnamefont
  {Bennett}}, \bibinfo {author} {\bibfnamefont {D.~P.}\ \bibnamefont
  {DiVincenzo}}, \bibinfo {author} {\bibfnamefont {T.}~\bibnamefont {Mor}},
  \bibinfo {author} {\bibfnamefont {P.~W.}\ \bibnamefont {Shor}}, \bibinfo
  {author} {\bibfnamefont {J.~A.}\ \bibnamefont {Smolin}}, \ and\ \bibinfo
  {author} {\bibfnamefont {B.~M.}\ \bibnamefont {Terhal}},\ }\href {\doibase
  10.1103/PhysRevLett.82.5385} {\bibfield  {journal} {\bibinfo  {journal}
  {Phys. Rev. Lett.}\ }\textbf {\bibinfo {volume} {82}},\ \bibinfo {pages}
  {5385} (\bibinfo {year} {1999}{\natexlab{b}})}\BibitemShut {NoStop}%
\bibitem [{\citenamefont {Shi}\ \emph {et~al.}(2020{\natexlab{b}})\citenamefont
  {Shi}, \citenamefont {Zhang},\ and\ \citenamefont
  {Chen}}]{PhysRevA.101.062329}%
  \BibitemOpen
  \bibfield  {author} {\bibinfo {author} {\bibfnamefont {F.}~\bibnamefont
  {Shi}}, \bibinfo {author} {\bibfnamefont {X.}~\bibnamefont {Zhang}}, \ and\
  \bibinfo {author} {\bibfnamefont {L.}~\bibnamefont {Chen}},\ }\href {\doibase
  10.1103/PhysRevA.101.062329} {\bibfield  {journal} {\bibinfo  {journal}
  {Phys. Rev. A}\ }\textbf {\bibinfo {volume} {101}},\ \bibinfo {pages}
  {062329} (\bibinfo {year} {2020}{\natexlab{b}})}\BibitemShut {NoStop}%
\bibitem [{\citenamefont {Bravyi}\ and\ \citenamefont
  {Smolin}(2011)}]{PhysRevA.84.042306}%
  \BibitemOpen
  \bibfield  {author} {\bibinfo {author} {\bibfnamefont {S.}~\bibnamefont
  {Bravyi}}\ and\ \bibinfo {author} {\bibfnamefont {J.~A.}\ \bibnamefont
  {Smolin}},\ }\href {\doibase 10.1103/PhysRevA.84.042306} {\bibfield
  {journal} {\bibinfo  {journal} {Phys. Rev. A}\ }\textbf {\bibinfo {volume}
  {84}},\ \bibinfo {pages} {042306} (\bibinfo {year} {2011})}\BibitemShut
  {NoStop}%
\bibitem [{\citenamefont {Wang}\ \emph {et~al.}(2014)\citenamefont {Wang},
  \citenamefont {Li},\ and\ \citenamefont {Fei}}]{PhysRevA.90.034301}%
  \BibitemOpen
  \bibfield  {author} {\bibinfo {author} {\bibfnamefont {Y.-L.}\ \bibnamefont
  {Wang}}, \bibinfo {author} {\bibfnamefont {M.-S.}\ \bibnamefont {Li}}, \ and\
  \bibinfo {author} {\bibfnamefont {S.-M.}\ \bibnamefont {Fei}},\ }\href
  {\doibase 10.1103/PhysRevA.90.034301} {\bibfield  {journal} {\bibinfo
  {journal} {Phys. Rev. A}\ }\textbf {\bibinfo {volume} {90}},\ \bibinfo
  {pages} {034301} (\bibinfo {year} {2014})}\BibitemShut {NoStop}%
\bibitem [{\citenamefont {Guo}(2016)}]{PhysRevA.94.052302}%
  \BibitemOpen
  \bibfield  {author} {\bibinfo {author} {\bibfnamefont {Y.}~\bibnamefont
  {Guo}},\ }\href {\doibase 10.1103/PhysRevA.94.052302} {\bibfield  {journal}
  {\bibinfo  {journal} {Phys. Rev. A}\ }\textbf {\bibinfo {volume} {94}},\
  \bibinfo {pages} {052302} (\bibinfo {year} {2016})}\BibitemShut {NoStop}%
\bibitem [{\citenamefont {Li}\ \emph {et~al.}(2014)\citenamefont {Li},
  \citenamefont {Wang},\ and\ \citenamefont {Zheng}}]{PhysRevA.89.062313}%
  \BibitemOpen
  \bibfield  {author} {\bibinfo {author} {\bibfnamefont {M.-S.}\ \bibnamefont
  {Li}}, \bibinfo {author} {\bibfnamefont {Y.-L.}\ \bibnamefont {Wang}}, \ and\
  \bibinfo {author} {\bibfnamefont {Z.-J.}\ \bibnamefont {Zheng}},\ }\href
  {\doibase 10.1103/PhysRevA.89.062313} {\bibfield  {journal} {\bibinfo
  {journal} {Phys. Rev. A}\ }\textbf {\bibinfo {volume} {89}},\ \bibinfo
  {pages} {062313} (\bibinfo {year} {2014})}\BibitemShut {NoStop}%
\bibitem [{\citenamefont {Chen}\ and\ \citenamefont
  {Dokovic}(2013)}]{Chen_2013}%
  \BibitemOpen
  \bibfield  {author} {\bibinfo {author} {\bibfnamefont {L.}~\bibnamefont
  {Chen}}\ and\ \bibinfo {author} {\bibfnamefont {D.~Z.}\ \bibnamefont
  {Dokovic}},\ }\href {\doibase 10.1088/1751-8113/46/27/275304} {\bibfield
  {journal} {\bibinfo  {journal} {Journal of Physics A: Mathematical and
  Theoretical}\ }\textbf {\bibinfo {volume} {46}},\ \bibinfo {pages} {275304}
  (\bibinfo {year} {2013})}\BibitemShut {NoStop}%
\bibitem [{\citenamefont {Hayashi}\ \emph {et~al.}(2006)\citenamefont
  {Hayashi}, \citenamefont {Markham}, \citenamefont {Murao}, \citenamefont
  {Owari},\ and\ \citenamefont {Virmani}}]{PhysRevLett.96.040501}%
  \BibitemOpen
  \bibfield  {author} {\bibinfo {author} {\bibfnamefont {M.}~\bibnamefont
  {Hayashi}}, \bibinfo {author} {\bibfnamefont {D.}~\bibnamefont {Markham}},
  \bibinfo {author} {\bibfnamefont {M.}~\bibnamefont {Murao}}, \bibinfo
  {author} {\bibfnamefont {M.}~\bibnamefont {Owari}}, \ and\ \bibinfo {author}
  {\bibfnamefont {S.}~\bibnamefont {Virmani}},\ }\href {\doibase
  10.1103/PhysRevLett.96.040501} {\bibfield  {journal} {\bibinfo  {journal}
  {Phys. Rev. Lett.}\ }\textbf {\bibinfo {volume} {96}},\ \bibinfo {pages}
  {040501} (\bibinfo {year} {2006})}\BibitemShut {NoStop}%
\bibitem [{\citenamefont {Bennett}\ \emph {et~al.}(1993)\citenamefont
  {Bennett}, \citenamefont {Brassard}, \citenamefont {Cr\'epeau}, \citenamefont
  {Jozsa}, \citenamefont {Peres},\ and\ \citenamefont
  {Wootters}}]{bennett1993teleporting}%
  \BibitemOpen
  \bibfield  {author} {\bibinfo {author} {\bibfnamefont {C.~H.}\ \bibnamefont
  {Bennett}}, \bibinfo {author} {\bibfnamefont {G.}~\bibnamefont {Brassard}},
  \bibinfo {author} {\bibfnamefont {C.}~\bibnamefont {Cr\'epeau}}, \bibinfo
  {author} {\bibfnamefont {R.}~\bibnamefont {Jozsa}}, \bibinfo {author}
  {\bibfnamefont {A.}~\bibnamefont {Peres}}, \ and\ \bibinfo {author}
  {\bibfnamefont {W.~K.}\ \bibnamefont {Wootters}},\ }\href {\doibase
  10.1103/PhysRevLett.70.1895} {\bibfield  {journal} {\bibinfo  {journal}
  {Phys. Rev. Lett.}\ }\textbf {\bibinfo {volume} {70}},\ \bibinfo {pages}
  {1895} (\bibinfo {year} {1993})}\BibitemShut {NoStop}%
\bibitem [{\citenamefont {Walgate}\ \emph {et~al.}(2000)\citenamefont
  {Walgate}, \citenamefont {Short}, \citenamefont {Hardy},\ and\ \citenamefont
  {Vedral}}]{PhysRevLett.85.4972}%
  \BibitemOpen
  \bibfield  {author} {\bibinfo {author} {\bibfnamefont {J.}~\bibnamefont
  {Walgate}}, \bibinfo {author} {\bibfnamefont {A.~J.}\ \bibnamefont {Short}},
  \bibinfo {author} {\bibfnamefont {L.}~\bibnamefont {Hardy}}, \ and\ \bibinfo
  {author} {\bibfnamefont {V.}~\bibnamefont {Vedral}},\ }\href {\doibase
  10.1103/PhysRevLett.85.4972} {\bibfield  {journal} {\bibinfo  {journal}
  {Phys. Rev. Lett.}\ }\textbf {\bibinfo {volume} {85}},\ \bibinfo {pages}
  {4972} (\bibinfo {year} {2000})}\BibitemShut {NoStop}%
\bibitem [{\citenamefont {Shamir}(1979)}]{shamir1979share}%
  \BibitemOpen
  \bibfield  {author} {\bibinfo {author} {\bibfnamefont {A.}~\bibnamefont
  {Shamir}},\ }\href {\doibase 10.1145/359168.359176} {\bibfield  {journal}
  {\bibinfo  {journal} {Commun. ACM}\ }\textbf {\bibinfo {volume} {22}},\
  \bibinfo {pages} {612–613} (\bibinfo {year} {1979})}\BibitemShut {NoStop}%
\end{thebibliography}%

\end{document}